\documentclass[12pt, draftclsnofoot, onecolumn]{IEEEtran}
\usepackage{caption}
\usepackage{setspace}
\usepackage{cite}
\ifCLASSINFOpdf
  \usepackage[pdftex]{graphicx}
\else
  \usepackage[dvips]{graphicx}
\fi
\usepackage{amsmath,amssymb,amsthm}
\usepackage{caption}
\usepackage{amsfonts}
\usepackage{algorithm,algorithmic}
\usepackage{array}
\usepackage{makecell}       
\usepackage{cite}
\usepackage{esint}
\usepackage{enumerate}
\usepackage{times}
\usepackage{url}
\usepackage{stfloats}
\usepackage{subfigure}
\usepackage{bbm}
\usepackage{multirow}
\ifCLASSOPTIONcompsoc
  \usepackage[caption=false,font=normalsize,labelfont=sf,textfont=sf]{subfig}
\else
  \usepackage[caption=false,font=footnotesize]{subfig}
\fi

\newtheorem{proposition}{\textbf{Proposition}}

\newcommand{\PreserveBackslash}[1]{\let\temp=\\#1\let\\=\temp}
\newcolumntype{C}[1]{>{\PreserveBackslash\centering}p{#1}}
\newcolumntype{R}[1]{>{\PreserveBackslash\raggedleft}p{#1}}
\newcolumntype{L}[1]{>{\PreserveBackslash\raggedright}p{#1}}

\begin{document}

  \title{Resource Consumption for Supporting Federated Learning in Wireless Networks}

 \author{
        Yi-Jing Liu,
        Shuang Qin\,~\IEEEmembership{Member,~IEEE},
        Yao Sun,~\IEEEmembership{Senior~Member,~IEEE},\\
        Gang~Feng,~\IEEEmembership{Senior~Member,~IEEE}
        


\thanks{

 Y. Liu, S. Qin, and G. Feng are with the National Key Lab on Communications, University of Electronic Science and Technology of China, Chengdu 611731, P. R. China, and also with the Yangtze Delta Region Institute (Huzhou), University of Electronic Science and Technology of China, Huzhou 313001, P. R. China. Y. Sun is with the James Watt School of Engineering, University of Glasgow, Glasgow G12 8QQ, U.K. S. Qin is the corresponding author (e-mail:blueqs@uestc.edu.cn).

}
  }%

\maketitle

\begin{abstract}
Federated learning (FL) has recently become one of the hottest focuses in wireless edge networks with the ever-increasing computing capability of user equipment (UE). In FL, UEs train local machine learning models and transmit them to an aggregator, where a global model is formed and then sent back to UEs. In wireless networks, local training and model transmission can be unsuccessful due to constrained computing resources, wireless channel impairments, bandwidth limitations, etc., which degrades FL performance in model accuracy and/or training time. Moreover, we need to quantify the benefits and cost of deploying edge intelligence, as model training and transmission consume certain amount of resources. Therefore, it is imperative to deeply understand the relationship between FL performance and multiple-dimensional resources. In this paper, we construct an analytical model to investigate the relationship between the FL model accuracy and consumed resources in FL empowered wireless edge networks. Based on the analytical model, we explicitly quantify the model accuracy, available computing resources and communication resources. Numerical results validate the effectiveness of our theoretical modeling and analysis, and demonstrate the trade-off between the communication and computing resources for achieving a certain model accuracy. 
\end{abstract}

\begin{IEEEkeywords}
Federated learning, edge intelligence, resource consumption, FL performance.
\end{IEEEkeywords}

\IEEEpeerreviewmaketitle

\section{Introduction}
Edge intelligence is boosted by the unprecedented computing capability of smart devices. Nowadays, more than 10 billion Internet-of-Things (IoT) equipment and 5 billion smartphones have emerged that are equipped with artificial intelligence (AI)-empowered computing modules, such as AI chips and graphic processing units (GPUs) \cite{IoT}. On the one hand, the user equipment (UE) can be potentially deployed as computing nodes to support emerging services, such as collaborative tasks, which paves the way for applying AI in wireless edge networks. On the other hand, in the paradigm of machine learning (ML), the powerful computing capability on these UEs can decouple conventional ML from acquiring, storing, and training data in data centers as conventional methods.

Federated learning (FL) has recently been widely acknowledged as one of the most essential enablers to bring edge intelligence into reality, as it facilitates collaborative training of ML models, while enhancing individual user privacy and data security \cite{lim2020federated,yang2019federated}. In FL, ML models are trained locally, therefore raw data remains in the device. Specifically, FL uses an iterative approach that requires a number of global iterations to achieve a certain global model accuracy. In each global iteration, UEs perform several local iterations to reach a local model accuracy \cite{lim2020federated,yang2019federated}. As a result, the implementation of FL in wireless networks can also reduce the costs of transmitting raw data, relieve the burden on backbone networks, and reduce latency for real-time decisions, etc.



While FL offers these attractive and valuable benefits, it also faces many challenges, especially when being deployed in wireless edge networks. For example, both local training and model transmission can be unsuccessful due to constrained resources and unstable transmission. Moreover, different from the conventional ML approaches, where raw datasets are sent to a central server, only the lightweight model parameters (\emph{i.e.}, weights, gradients, etc.) are exchanged in FL. Nevertheless, the communication cost of FL could be still fairly large and cannot be ignored. The experimental results in \cite{jeong2018communication} show that the model size of a 5-layer convolutional neural network used for MNIST (classification) is about 4.567MB per global iteration for $28\times28$ images, while the model size of ResNet-110 used for CIFAR-10 (classification) is around 4.6MB per global iteration for $32\times32$ images \cite{he2020group}. Therefore, before deploying FL empowered wireless edge networks, we need to answer two fundamental questions: (1) How accurate of an ML model can be achieved by using FL, and (2) How much cost is incurred to guarantee certain required FL performance? Obviously, answering these two questions is of paramount importance for facilitating edge network intelligence. Therefore, we need to deeply understand the relationship between FL performance and consumed multi-dimensional resources.

In this paper, we theoretically analyze how many resources are needed to support an FL-empowered wireless edge network by assuming spatial-temporal domain Poisson distribution. We first derive the distribution of signal-to-interference-plus-noise ratio (SINR), signal-noise ratio (SNR), model transmission success probability, and resource consumption. Then, we evaluate the impact of the amount of resources on FL performance. Numerical results validate the accuracy of our theoretical modeling and analysis. The main contributions of this paper can be summarized as follows,
\begin{enumerate}
\def\labelenumi{(\arabic{enumi})}
\item
We develop an analytical model for FL empowered wireless edge networks, where UE geographical distribution and arrival rate of the interfering UEs are modeled as Poisson Point Process (PPP).

\item
We theoretically analyze SINR, SNR, and the local/global model transmission success probability. Specifically, we derive the probability density function (PDF) of SINR and SNR, where we obtain the transmission success probability of the local/global model. 
\item
Based on the analytical model, we derive the explicit expression of the model accuracy, as a function of the amount of resources (including communication resources and computing resources) under FL framework. 
\item
We investigate three specific cases according to the sufficiency of respective communication and computing resources. We use simulation experiments to validate the effectiveness of our theoretical modeling and analysis, and demonstrate the trade-off relationship between the communication resources and computing resources for achieving certain machine learning model accuracy.


\end{enumerate}

In the rest of this paper, we review related work in Section II. Then we present the FL empowered edge network model in Section III and the analysis for the communication and computing resource consumption in Section IV. In Section V, the relationship between FL performance and consumed resources is derived. In Section VI, different cases based on the sufficiency of respective communication and computing resources are discussed. Finally, we present the numerical results in Section VII and conclude the paper in Section VIII.

\section{Related Work}

Currently, there has been a large body of work on developing various FL algorithms for FL empowered wireless edge networks. The authors of \cite{chen2020convergence} designed an appropriate user selection scheme to minimize FL convergence time and training loss by jointly optimizing user selection and resource block allocation. The authors of \cite{chen2020wireless} proposed a collaborative FL framework to enable UEs to implement FL with less reliance on a central server by aggregating the local FL models received from the associated UEs. In \cite{9247530}, the authors presented a Stackelberg-game-based approach to develop an FL incentive scheme by modeling the incentive-based interaction between a global server and participating UEs. In \cite{liu2020device}, we proposed a hybrid FL scheme to make a global UE association decision for heterogeneous models by exploiting two levels of model aggregation. All aforementioned investigations aimed to facilitate edge intelligence by developing suitable FL algorithms in wireless edge networks. However, the authors have not explicitly addressed the resource cost under FL framework. In fact, the communication cost under FL framework could be still fairly large and cannot be ignored, although only the lightweight model parameters are exchanged.

To support these improved FL algorithms even the legacy FL algorithms, resource-efficient and FL performance guarantee are indispensable basis for achieving the FL empowered wireless edge intelligence. The authors of \cite{9488679} developed a low-cost sampling-based algorithm by adapting various control variables to minimize cost components (\emph{e.g.}, learning time, energy consumption). In \cite{9488679}, the authors considered a multivariate control problem for energy-efficient FL to guarantee convergence by designing principles for different optimization goals. The authors of \cite{8952884} proposed an over-the-air computation based approach to improve communication efficiency by modeling joint device selection and beamforming design as a sparse and low-rank optimization problem. In \cite{9430906}, the authors introduced update-importance-based client scheduling schemes to reduce the required number of model training rounds by selecting a subset of clients for local updates in each round of training. The authors of \cite{9322580} proposed a convergent over-the-air FL scheme to reduce bandwidth and energy consumption by inducing precoding and scaling upon transmissions to gradually mitigate the effect of the noisy channel. In \cite{wen2021federated}, the authors proposed a federated dropout scheme to enable FL on resource-constrained devices by tackling both the communication and computation resource bottlenecks. All these investigations aimed to achieve resource-efficient FL algorithms in wireless edge networks. However, the vulnerability of wireless links is largely ignored, which directly degrades FL performance by affecting local training and model transmission. Therefore, deeply understanding the relationship among FL performance, wireless factors, and multi-dimensional resources is essential for enabling wireless edge intelligence.

So far, there has been little attention on quantifying the relationship between FL performance and consumed resources, while considering the vulnerability of wireless links. The authors of \cite{LAC} investigated the trade-off between the number of local iterations and the number of global iterations to capture the relationship between FL training time and energy consumption. However, they have not considered the unsuccessful model transmission in a real network. In \cite{Ajoint}, the authors studied a joint learning and communication optimization scheme to minimize an FL loss function, where the limited resources and unstable wireless links were considered. However, they only focused on optimizing FL performance without comprehensively quantifying the relationship between FL performance and consumed resources.

\section{FL empowered Wireless Network Model}
We consider an FL empowered wireless edge network consisting of a central base station (BS) and multiple UEs, as shown in Fig.1. The UEs can be regarded as local computing nodes for local model training, while the server (\emph{e.g.}, edge servers) serves as the model aggregator co-located with the BS \cite{lim2020federated,liu2020device,yang2019federated}. To quantitatively present the FL empowered wireless network model, we need to model the distribution of the UEs and arrival rate of the interfering UEs. As one of the most commonly used point processes, PPP model has been widely used to model UEs distribution and/or arrival rate of the interfering UEs in wireless networks, where a huge amount of data has validated the accuracy of the model \cite{flint2017analysis,hunter2008transmission,weber2010overview,SunBC}. Nevertheless, other point processes, such as Poisson cluster process (PCP) \cite{7110502} and cox process \cite{8340239} for some specific scenarios, can also be used in our analytical model.
\begin{figure}[!htbp]
\vspace{-5mm}
\centering
\includegraphics[width=5cm]{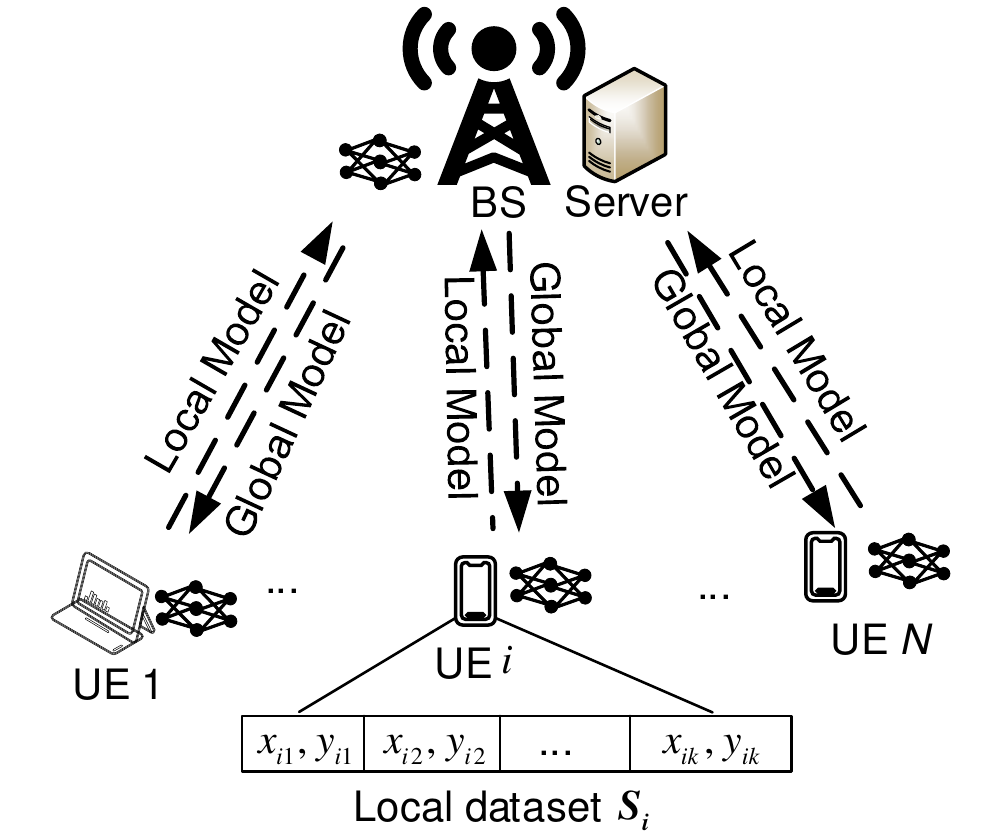}
\caption{The FL empowered wireless edge network.}
\label{fig:1}
\vspace{-7mm}
\end{figure}

\subsection{FL Model}
\subsubsection{Loss Function}
Let random variable $N$ denote the number of UEs that are geographically distributed as homogeneous PPP with intensity $\lambda_i$, where $n$ denotes the value of $N$. Similarly, we use capital letter to denote a random variable and the corresponding lower case to its value in the following. For convenience, the frequently used notations are summarized in TABLE I.

\setcounter{table}{0}
\renewcommand{\tablename}{TABLE} 
\renewcommand{\thetable}{\Roman{table}} 
\begin{table*}[htbp] \label{tab:FREQUENTLY USED NOTATIONS}
\small
\vspace*{-2mm}
\centering
\caption{List of notations}
\setlength{\tabcolsep}{1mm}{
\begin{tabular}{cc||cc}
\hline
Notation & Definition & Notation & Definition\\\hline
$N$ & Number of UEs (random variable) & $n$ & The value of $N$\\
$N_I$ & Number of interfering UEs & $n_I$ & The value of $N_I$\\ 
$N_{\cal A}$& Number of UEs in interfering area $\cal A$& $n_{\cal A}$ & The value of $N_{\cal A}$\\
$\lambda_i$ & UE density & $\lambda_a$ & Interfering UE arrival rate\\ 
${\cal S}_i$ &  Datasets of UE $i$ & ${S}_i$ & The amount of ${\cal S}_i$\\
$t$ & The index of local iterations, $0\leq t\leq \tau$&$s_i$ & The value of $S_i$\\
$\tau$ & Number of local iterations&$K$ & Number of communication rounds\\
$w_r$ & Global model at $r$-th round & $w_i^r(t)$ & Local model of UE $i$\\
$Z_i$& Computing capacity of UE $i$& $I_i$& Interference of UE $i$\\
$P_{\text{up}}$ & Transmit power of the UE & $P_{\text{down}}$ & Transmit power of the BS \\ 
$D_1$&Distance between the UE and the BS& $r_0$ & Radius of the BS coverage\\
${\bf D}_2$ & Distance vector for interfering UEs& $d_0$ & Radius of interfering area\\
$\mathit{SINR}_{\text{up}}$& SINR of uplink &$\mathit{SNR}_{\text{down}}$& SNR of downlink\\
$\beta_{\text{down}}$ & SNR threshold &$\beta_{\text{up}}$ & SINR threshold \\
$R_{\text{down}}^{i,r}$&Transmission rate (downlink)&$R_{\text{up}}^{i,r}$&Transmission rate (uplink)\\
$b_{\text{down}}^{i,r}$ &Bandwidth consumption (downlink)&$b_{\text{up}}^{i,r}$ & Bandwidth consumption (uplink)\\
$T_i$ & Training time of UE $i$ for  & $c_i$&Number of CPU cycles for \\
& one local iteration & & computing a sample\\

\hline
\end{tabular}
}
\vspace*{-6mm}
\end{table*}





For a specific UE $i$, it has a local dataset ${\cal S}_i$ with $S_i$ data samples, where ${\cal S}_i=\{x_{ik} \in \mathbb{R}^d, y_{ik} \in \mathbb{R}\}_{k=1}^{S_i}$. Moreover, we define $f_k(w_i^r(t);x_{ik}, y_{ik})$ as a loss function for data sample $k$ of UE $i$, where $w_i^r(t)$ represents the model parameter of UE $i$ at the $t$-th local iteration during the $r$-th global iteration. The loss function $f_k(w_i^r(t);x_{ik}, y_{ik})$ is different for various FL learning tasks \cite{hennig2007some}. For example, for a linear regression, the loss function is $f_k(w_i^r(t);x_{ik}, y_{ik})=\frac{1}{2}(x_{ik}^{\text{T}}w_i^r(t)-y_{ik})^2$. For neural network, the loss function could be mean squared error (\emph{i.e.}, $\frac{1}{n}\sum_{i=1}^{n}(y_{ik}-\widehat{y}_{ik})$), where $\widehat{y}_{ik}$ represents the predicted value of $y_{ik}$. Based on $f_k(w_i^r(t);x_{ik}, y_{ik})$, we define $F_i(w_i^r(t)): \mathbb{R}^m\rightarrow \mathbb{R}$ as a local loss function to capture the local training performance, which is as follows,
\begin{equation}
\setlength{\abovedisplayskip}{2pt}
\setlength{\belowdisplayskip}{2pt}
F_i(w_i^r(t))\triangleq \frac{1}{S_i}\sum_{k\in S_i}f_k(w_i^r(t);x_{ik}, y_{ik}).
\end{equation}

In addition, we define $F(w_r)$ as the global loss function on all distributed datasets to measure the global training performance, which is expressed by
\begin{equation}\label{eq:2}
\setlength{\abovedisplayskip}{2pt}
\setlength{\belowdisplayskip}{2pt}
F(w_r)\triangleq \frac{\sum_{i=1}^n S_i F_i(w_i^r(t))}{S}, 
\end{equation}
where ${S}\triangleq \sum_{i=1}^n {S}_i$. The goal of the BS is to derive a vector $w_r^*$ satisfying $w_r^{*}\triangleq\arg_{w_r}\min F(w_r)$. 

\subsubsection{Updating Model}
In FL, each global iteration is called a \emph{communication round} \cite{liu2020device}, as shown in Fig.2. A communication round consists of $5$ phases including local model updating, local iterations (also called local training), local model transmission, global model updating, and global model transmission. In the following, we present the details of the local and global model updating respectively.
\begin{figure}[!htbp]
\vspace{-5mm}
\centering
\includegraphics[width=7cm]{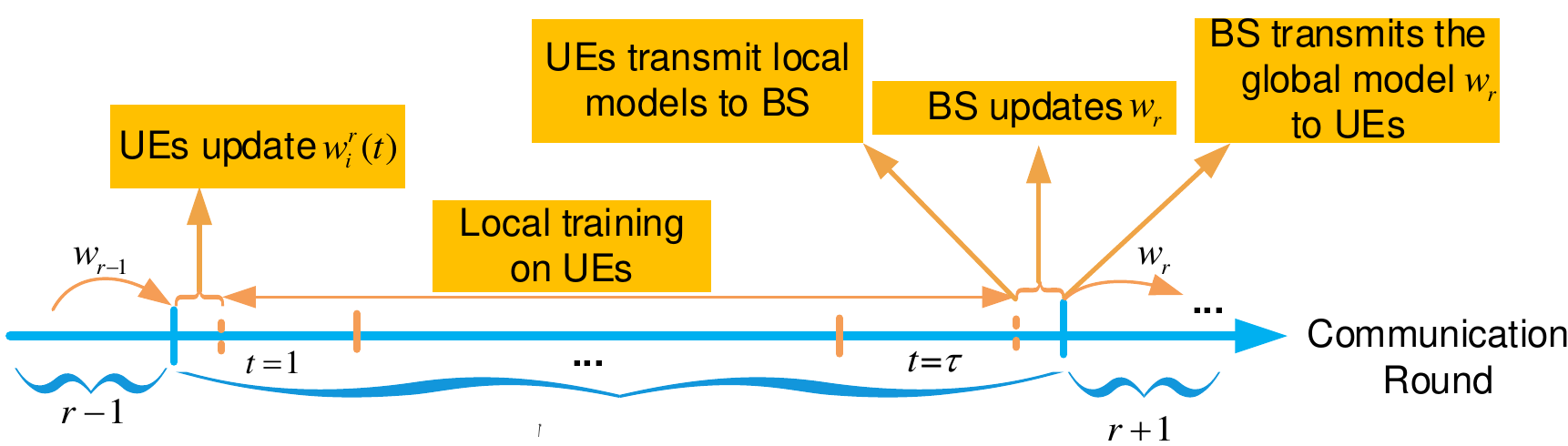}
\caption{The process of a communication round.}
\label{fig:2}
\vspace{-5mm}
\end{figure}
\begin{enumerate}
\def\labelenumi{(\arabic{enumi})}
\item
\emph{Local Model Updating}: The local model updating can be performed based on a local learning algorithm, such as gradient descent (GD), actor-critic (AC), etc. Specifically, if $0<t\leq \tau$, the local model $w_i^r(t)$ is updated by $w_i^r(t)=w_{r}-\eta\nabla F_i(w_i^r(t-1))$, while $w_i^r(t)=w_{r}$ if $t=0$,
where $t$ represents the index of local iterations, $\tau$ represents the total number of local iterations during a communication round. Moreover, $\eta \geq 0$ is the step size, and $w_r$ represents the global model at the $r$-th communication round.
\item
\emph{Global Model Updating}: After $\tau$ local iterations, \emph{i.e.}, $t=\tau$, UEs will achieve a certain local accuracy and send the local models $w_i^r(t)$ to the aggregator. Then the global aggregation is performed at the aggregator according to
\begin{equation}\label{eq:5}
\setlength{\abovedisplayskip}{2pt}
\setlength{\belowdisplayskip}{2pt}
w_r=\frac{\sum_{i=1}^n {S}_i w_i^r(t)}{S}, t=\tau.
\end{equation}
\end{enumerate}

\subsection{Computing Resource Consumption Model}
For a specific UE $i$, let ${Z}_i$ denote its computing capacity in cycles/s. $c_i$ (cycles/sample) denotes the number of CPU cycles required for computing one data sample at UE $i$. $T_i$ represents the local computing time (training time) needed for one local iteration. Therefore, similar to that in \cite{yang2020energy}, the consumed computing resources during one local iteration for UE $i$ is given by ${Z}_i=\frac{c_i{S}_i}{T_i}$.



\subsection{Communication Resource Consumption Model}
\subsubsection{Uplink}
The transmission time for UE $i$ to transmit the local model $w_i^r(t)$ (uplink direction) is denoted by ${T}_\text{up}^{i,r}$. Since the dimensions of local models are fixed for all UEs that participate in local training, the data size of the local model on each UE is constant and is denoted by $s$ \cite{yang2020energy}. The transmission rate of UE $i$ on the wireless channel to the BS during the $r$-th communication round is represented by $R_{\text{up}}^{i,r}$. Therefore, we have $\frac{s}{R_{\text{up}}^{i,r}}= {T}_\text{up}^{i,r}$, where
\begin{equation}\label{eq:Rup}
\setlength{\abovedisplayskip}{2pt}
\setlength{\belowdisplayskip}{2pt}
R_{\text{up}}^{i,r}=b_{\text{up}}^{i,r}\log_2(1+\mathit{SINR}_{\text{up}}(D_1, N_I, \mathbf{D}_2)), 
\end{equation}
where $b_{\text{up}}^{i,r}$ represents the amount of consumed bandwidth for transmitting the local model $w_i^r(t)$. In addition, $\mathit{SINR}_{\text{up}}(D_1, N_I, \mathbf{D}_2)=\frac{P_{\text{up}}G(D_1)}{\sum_{j=1}^{N_I}PG({D}_2^{(j)})+\delta^2}$ is the SINR. $D_1$ represents the distance between the UE and the BS, $\mathbf{D}_2=[{D}_2^{(1)}, {D}_2^{(2)},...{D}_2^{(j)},...,{D}_2^{(N_I)}]$ denotes the distance vector for all interfering UEs of UE $i$, $N_I$ represents the number of interfering UEs with $N_I\leq N$, $\delta^2$ denotes the noise power, $P_{\text{up}}$ represents the transmit power of the UE. $G(\cdot)$ represents the large-scale channel gain between the BS and UEs. Indeed, the channel gain model could be either large-scale (\emph{e.g.}, path loss) or small-scale (\emph{e.g.}, Rayleigh fading, Rician fading), which only affects SINR/SNR. Furthermore, let $\beta_{\text{up}}$ denote the SINR threshold that the BS can successfully decode the received updates from UE $i$. Therefore, local model transmission is successful only if $\mathit{SINR}_{\text{up}}(D_1, N_I, \mathbf{D}_2)>\beta_{\text{up}}$. 


\subsubsection{DownLink}
Let us analyze SNR for the BS to transmit the global model (downlink direction). The transmission time for transmitting the global model $w_r$ in the downlink is denoted by ${T}_{\text{down}}^{i,r}$. From equation (\ref{eq:5}), we see that the dimensions of the global model $w_r$ is similar to that of local models. Therefore, the data size of the global model that the BS sends to each UE is also equal to $s$ \cite{Ajoint}. We assume that the transmission rate of the BS during the $r$-th communication round is represented by $R_{\text{down}}^{i,r}$. Therefore, we have $\frac{s}{R_{\text{down}}^{i,r}}={T}_{\text{down}}^{i,r}$, where
\begin{equation}\label{eq:Rdown}
\setlength{\abovedisplayskip}{2pt}
\setlength{\belowdisplayskip}{2pt}
R_{\text{down}}^{i,r}=b_{\text{down}}^{i,r}\log_2(1+\mathit{SNR}_{\text{down}}(D_1)),
\end{equation}
where $b_{\text{down}}^{i,r}$ represents the consumed bandwidth for transmitting the global model. In addition, $\mathit{SNR}_{\text{down}}(D_1)=\frac{P_{\text{down}}G(D_1)}{\delta^2}$, where $P_{\text{down}}$ represents the transmit power of the BS allocated to all UEs. Furthermore, let $\beta_{\text{down}}$ denote the SNR threshold that UEs can successfully decode the received updates from the BS. Therefore, the global model transmission is successful only if $\mathit{SNR}_{\text{down}}(D_1)> \beta_{\text{down}}$. 

\section{Wireless Bandwidth and Computing Resources Consumed for Supporting FL empowered Edge Intelligence}
A certain amount of wireless bandwidth is consumed in the uplink/downlink, when models are exchanged between the BS and the UEs. Specifically, in the uplink, the UEs send their local models to the BS via some channel partitioning schemes, such as orthogonal frequency division multiplexing (OFDM). in the downlink, the BS sends the global model to individual UEs. Indeed, the wireless transmission environment of the uplink/downlink will affect the transmission process of the local/global model, and thus affect the global aggregation and local training. In this section, we theoretically analyze SINR, SNR, as well as wireless bandwidth consumed in the uplink/downlink to support FL empowered wireless edge networks.



\subsection{SINR Analysis for Uplink}
\subsubsection{Probability Density Function (PDF) of SINR} 
To derive the PDF of SINR, we separately investigate the signal power and interference. We have assumed that the UEs are geographically distributed as homogeneous PPP with intensity $\lambda_i$, and thus the number of UEs $N$ is a variable of Poisson distribution with density parameter $\pi(r_0)^2\lambda_i$, where $r_0$ represents the radius of the BS coverage. For a specific UE $i$, the signal power is also a random variable, \emph{i.e.}, $S_{\text{up}}=P_{\text{up}}G(D_1)$, as it only relates to the distance $D_1$ and $P_{\text{up}}$ is always fixed for each UE. 
\begin{proposition} \notag \label{Prop1}
\it{The PDF of the distance $D_1$ between a specific UE and the serving BS is $f_{D_1}(d_1)=\frac{2d_1}{(r_0)^2}$.} 
\end{proposition} 
\begin{IEEEproof}
As the PDF of location $(X,Y)$ for UE $i$ is $f(X,Y)=\frac{1}{\pi (r_0)^2}$, the CDF of distance $D_1=d_1$ is $F_{D_1}(d_1)=\iint \limits_{X^2+Y^2 \leq (d_1^2)} \frac{1}{\pi (r_0)^2} dXdY=\frac{(d_1)^2}{\left(r_0\right)^2}$. Therefore, the PDF of $D_1=d_1$ is $f_{D_1}(d_1)=F^{'}_{D_1}(d_1)=\frac{2d_1}{(r_0)^2}$.
\end{IEEEproof}

Therefore, we can obtain the PDF of signal power (\emph{i.e.}, $f_S (S=P_{\text{up}}G(D_1))=f_{D_1}(d_1)$) for deriving the closed-form expression of ${\rm Pr}(\mathit{SINR}_{\text{up}}>\beta_{\text{up}})$. Next, we investigate the distribution of the received interference in the uplink. Note that only the transmitting UEs located in the interfering area with radius $d_0$, can contribute to the interference. We assume that 
the number of UEs within the interfering area is represented by $N_{\cal A} (N_{\cal A}\geq N_I)$, which is also a variable of Poisson distribution with density parameter $\pi (d_0)^2 \lambda_i$. Moreover, the transmission time for the UEs is represented by ${t}_{\text{up}}$, where the transmitting UEs during time $[-t_{\text{up}}, t_{\text{up}}]$ can contribute to interference. Therefore, for a specific UE, the number of interfering UEs is distributed as PPP with parameter $2t_{\text{up}}\lambda_a$, where $\lambda_a$ denotes the arrival rate of interfering UEs. Therefore, the interference probability of a transmitting UE during time $[-t_{\text{up}},t_{\text{up}}]$ is ${\rm Pr}(active)=1-\text{exp}\{-2t_{\text{up}}\lambda_a\}$.

Therefore, the probability of the number of interfering UEs $N_I=n_I$ given $N_{\cal A}=n_{\cal A}$ is 
\begin{equation}
\setlength{\abovedisplayskip}{1pt}
\setlength{\belowdisplayskip}{1pt}
{\rm Pr}(N_I=n_I|N_{\cal A}=n_{\cal A})=C_{n_{\cal A}}^{n_I}(1-\text{exp}\{1-2t_{\text{up}}\lambda_a\})^{n_I}\cdot
(\text{exp}\{1-2t_{\text{up}}\lambda_a\})^{n_{\cal A}-n_I},
\end{equation}
where $C_{n_{\cal A}}^{n_I}$ is the combination number. Therefore, the PDF of $N_I$ is
\begin{equation}\label{eq:17}
\setlength{\abovedisplayskip}{1pt}
\setlength{\belowdisplayskip}{1pt}
f_{N_I}(n_I)={\rm Pr}(N_I=n_I)
=\sum_{n_{\cal A}=n_I}^N{\rm Pr}(N_I=n_I|N_{\cal A}=n_{\cal A}){\rm Pr}(N_{\cal A}=n_{\cal A}),
\end{equation}
where ${\rm Pr}(N_{\cal A}=n_{\cal A})=\frac{(\pi (d_0)^2 \lambda_i)^{n_{\cal A}}}{n_{\cal A}!} \text{exp}\{-\pi (d_0)^2 \lambda_i\}$. Based on Proposition 1, we can derive the PDF of interference $I_i$ generated by UE $i$, \emph{i.e.}, $f_{I_i}(I_i=P_{\text{up}}G(d_2^{(i)}))=f_{D_2^{(i)}}(d_2^{(i)})=\frac{2d_2^{(i)}}{d_0^2}$. As the total interference $I(N_I, \mathbf{D}_2)$ is affected by the number of interfering UEs $N_I$ as well as the distance of these interfering UEs $\mathbf{D}_2$, we have the PDF of $I(N_I, \mathbf{D}_2)$, as follows,
\begin{equation}\label{eq:25}
\setlength{\abovedisplayskip}{1pt}
\setlength{\belowdisplayskip}{1pt}
f_I(N_I=n_I, \mathbf{D}_2={\mathbf{d}}_2)=f_{N_I}(n_I)Pr(\mathbf{D}_2=\mathbf{d}_2|N_I=n_I)
=f_{N_I}(n_I)\left(\frac{2}{(d_0)^{2}}\right)^{n_I}\prod^{n_I}_{n=1} d_2^{(n)}.~~
\end{equation}

Therefore, the PDF of SINR can be given by
\begin{equation}
\setlength{\abovedisplayskip}{1pt}
\setlength{\belowdisplayskip}{1pt}
f_{\mathit{SINR}_{\text{up}}}(D_1=d_1, N_I=n_I, \mathbf{D}_2={\mathbf{d}}_2)
=f_{D_1}(d_1)f_I(N_I=n_I, \mathbf{D}_2={\mathbf{d}}_2),
\end{equation}

\subsubsection{transmission success probability of Local Models}
Local model transmission is successful if $\mathit{SINR}_{\text{up}}> \beta_{\text{up}}$. Therefore, the transmission success probability of local models is given by 
\begin{equation}\label{eq:threshold}
\setlength{\abovedisplayskip}{1pt}
\setlength{\belowdisplayskip}{1pt}
{\rm Pr}(\mathit{SINR}_{\text{up}}>\beta_{\text{up}})=\iiint \limits_{\cal A} f_{\mathit{SINR}_\text{up}}d{\cal A},
\end{equation}
where ${\cal A}$ represents the area of $(D_1, N_I, \mathbf{D}_2)$ that satisfies $\mathit{SINR}_{\text{up}}(D_1, N_I, \mathbf{D}_2)> \beta_{\text{up}}$. As we can obtain $f_{\mathit{SINR}_\text{up}}$ in equation (\ref{eq:threshold}), we only need to find the interfering area ${\cal A}$. 

For the distance between the UE and the serving BS, intuitively $f_{\mathit{SINR}_\text{up}}<\beta_{\text{up}}$, when $D_1>d_0$ \cite{SunBC}. Therefore, the satisfying range of $D_1$ is $(0, d_0]$. Therefore, when given $D_1=d_1$, we can obtain the number of interfering UEs $N_I$ and the location of these interfering UEs. Let $\overline{n}_I$ represent the mean of random variable $N_I$. Based on the UE distribution and interfering UE arrival models, we can derive $\overline{n}_I$ as follows,
\begin{equation}
\setlength{\abovedisplayskip}{1pt}
\setlength{\belowdisplayskip}{1pt}
\overline{n}_I\triangleq E(N_{\cal A}){\rm Pr}(active)=\pi (d_0)^2\lambda_i\left(1-\text{exp}\{1-2t_{\text{up}}\lambda_I\}\right).
\end{equation}

Therefore, SINR is only related to $D_1$ and $\mathbf{D}_2$, expressed as $\mathit{SINR}_{\text{up}}(D_1, \mathbf{D}_2)=\frac{P_{\text{up}}G(D_1)}{\sum_{i=1}^{\overline{n}_I}I_i+\delta^2}$,
where $I_i=P_{\text{up}}G(D_2^{(i)})$ represents the interference generated by UE $i$. Therefore, we have 
\begin{equation}
\setlength{\abovedisplayskip}{1pt}
\setlength{\belowdisplayskip}{1pt}
{\rm Pr}(\mathit{SINR}_{\text{up}}>\beta_{\text{up}})
={\rm Pr}\left(\sum_{i=1}^{\overline{n}_I}I_i<\frac{P_{\text{up}}G(D_1)}{\beta_{\text{up}}}-\delta^2\right).
\end{equation}

In a typical FL framework, the number of UEs involved in local model training is fairly large, say at least hundreds of UEs \cite{wang2021field}. Therefore, based on the central limit theorem, $\sum_{i=1}^{\overline{n}_I}I_i$ follows a \emph{normal distribution} $N(\mu_I, \sigma_I^2)$ \cite{SunBC,Hsu1947}. Furthermore, we have $\mu_I=\overline{n}_IE(I_i)$ and $\sigma_I=\sqrt{\overline{n}_I}D(I_i)$, which are the mean and variance of $I_i$ respectively \cite{Hsu1947}. More details about $\mu_I$ and $\sigma_I$ can be found in Appendix A.

Let $Y=\frac{I-\mu_I}{\sigma_I}$, where $I=\sum_{i=1}^{\overline{n}_I}I_i$ and $I\sim N(\mu_I, \sigma_I)$. Therefore, we have $Y\sim N(0,1)$, where
\begin{equation}
\setlength{\abovedisplayskip}{1pt}
\setlength{\belowdisplayskip}{1pt}
{\rm Pr}\left(\sum_{i=1}^{\overline{n}_I}I_i<\frac{P_{\text{up}}G(d_1)}{\beta_{\text{up}}}-\delta^2\right)
={\rm Pr} \left(Y<\frac{\frac{P_{\text{up}}G(d_1)}{\beta_{\text{up}}}-\delta^2-\mu_I}{\sigma_I}\right)=\Phi(\xi(d_1)),
\end{equation}
where $\xi(d_1)=\frac{\frac{P_{\text{up}}G(d_1)}{\beta_{\text{up}}}-\delta^2-\mu_I}{\sigma_I}$ and $\Phi$ represents the cumulative distribution function (CDF) of standard normal distribution. Therefore, we have
\begin{equation}\label{eq:SINR}
\setlength{\abovedisplayskip}{2pt}
\setlength{\belowdisplayskip}{1pt}
{\rm Pr}\left(\mathit{SINR}_{\text{up}}>\beta_{\text{up}}\right)=\iiint \limits_{\cal A} f_{\mathit{SINR}_{\text{up}}} d{\cal A}=\int_{d_1=d_\text{min}}^{d_0} f_{D_1}(d_1) \Phi(\xi(d_1))d(d_1).
\end{equation}




\subsection{SNR Analysis for Downlink}
As $\text{SNR}_{\text{down}}(D_1)=P_{\text{down}}G(D_1)/\delta^2$, we can obtain the PDF of the signal power, as $f_{S_{\text{down}}}(S_{\text{down}}=P_{\text{down}}G(D_1))=f_{D_1}(d_1)$, when given transmit power $P_{\text{down}}$ and noise level $\delta^2$. Therefore, given $G(d_\text{min}^{'})=\delta^2\beta_{\text{down}}/P_{\text{down}}$, where we assume $G(d_1)$ monotonically increases, we have
\begin{equation}\label{eq:SNR}
\setlength{\abovedisplayskip}{2pt}
\setlength{\belowdisplayskip}{1pt}
{\rm Pr}(\mathit{SNR}_{down}>\beta_{\text{down}})={\rm Pr}(\frac{P_{\text{down}}G(D_1)}{\delta^2}>\beta_{\text{down}})
=\int_{d_1=d_\text{min}^{'}}^{r_0} f_{D_1}(d_1)d(d_1).
\end{equation}

\subsection{Wireless Bandwidth Consumed for Transmitting Local/Global Models}
Based on equation (4), the bandwidth consumed for transmitting the local model $w_i^r(t)$ during the $r$-th communication round is given by $b_{\text{up}}^{i,r}=\frac{s}{T_{\text{up}}^{i,r}\log_2(1+\mathit{SINR}_{\text{up}}(D_1,N_I,\textbf{D}_2)}$. 
As $s$ and $T_{\text{up}}^{i,r}$ are constant, the PDF of $b_{\text{up}}^{i,r}$ for UE $i$ in the uplink is equal to $f_{\mathit{SINR}_{\text{up}}}$. Therefore, the mean of bandwidth for all UEs transmitting local models during $K$ communication rounds is as follows,
\begin{equation}\label{eq:BR}
\setlength{\abovedisplayskip}{1pt}
\setlength{\belowdisplayskip}{1pt}
\overline{B}_{\text{up}}=K\cdot \sum_{i=1}^{n} b_{\text{up}}^{i,r} f_{\mathit{SINR}_{\text{up}}}.
\end{equation}
Similarly, the mean of bandwidth for transmitting the global models during $K$ communication rounds is given by
\begin{equation}\label{eq:BRD}
\setlength{\abovedisplayskip}{-1pt}
\setlength{\belowdisplayskip}{-1pt}
\overline{B}_{\text{down}}=K \cdot \sum_{i=1}^{n} b_{\text{down}}^{i,r} f_{\mathit{SNR}_{\text{down}}}.
\end{equation}


\vspace{-5mm}
\subsection{Consumed Computing Resources in FL}
We assume the processing capacity of all UEs are constant in cycles/sample. Moreover, as different ML models pose different degrees of complexity, we assume all UEs train the same FL task in our analytical model, where the local ML models have the same size and structure. Therefore, the total amount of computing resources needed to support local model training for all UEs is affected by the number of training UEs as well as the amount of datasets on the UEs. On the one hand, the number of training UEs is affected by the wireless transmission of the global model. On the other hand, many of the existing studies explicitly indicate that the amount of different datasets distributed on the UEs is imbalanced, as the data is collected directly and stored persistently \cite{DBLP,pmlr-v54-mcmahan17a}. Note here data imbalance means the different amount of local datasets, instead of different dataset contents, so our analytical model is based on i.i.d. data, where non-i.i.d data case is left for future work. Therefore, we theoretically analyze computing resources consumption for supporting local training from the perspective of SNR and imbalanced datasets in this section.


We assume that the amount of datasets on the UEs follows the \emph{normal distribution} \cite{gao2020end,liu2020deep}, \emph{i.e.}, ${\cal S}_i\sim N(\mu_i, \sigma_i^2)$, where $\mu_i$ or/and $\sigma_i^2$ could be different for specific UEs. Indeed, other distributions, such as Beta distribution and Gamma distribution, can also be used in our analytical model. Moreover, as the computing resources consumption of UE $i$ for one local iteration is ${Z}_i=\frac{c_iS_i}{T_i}$, the PDF of ${Z}_i$ is equal to $f_{{S}_i}(s_i)$, \emph{i.e.}, $f_{{Z}_i}(z_i)=\frac{1}{\sqrt{2\pi}\sigma_{i}}\text{exp}\left(-\frac{(s_i-\mu_{i})^2}{2\sigma_{i}^2}\right)$.



For a specific UE $i$, if $\mathit{SNR}_{down}>\beta_{\text{down}}$, we say UE $i$ can successfully receive the global model. In other words, UE $i$ will continue to perform local training in the next communication round and consume certain computing resources. Let $\boldsymbol{\hat{Z}}=\{\hat{Z}_1, \hat{Z}_2,...,\hat{Z}_i,...,\hat{Z}_n\}$ indicate the certain computing resources consumed by all UEs, where the value of $\hat{ Z}_i$ is set to ${z}_i$ if UE $i$ successfully receives the global model and 0 otherwise. Therefore, we can obtain the PDF of $\hat{Z}_i$ as follows,
\begin{equation}\label{eq:C_i}
\setlength{\abovedisplayskip}{1pt}
\setlength{\belowdisplayskip}{1pt}
f_{\hat{Z}_i}(\hat{Z}_i=z_i)={\rm Pr}(\hat{Z}_i=z_i|\mathit{SNR}_{down}>\beta_{\text{down}})=f_{{Z}_i}(z_i){\rm Pr}(\mathit{SNR}_{down}>\beta_{\text{down}}).
\end{equation}

Therefore, based on equation (\ref{eq:C_i}), we can derive the mean of computing resources consumed by all UEs for one local iteration, which is given by $\overline{C}_{\mathit{UE}}=\sum_{i=1}^{n}z_if_{\hat{Z}_i}(\hat{Z}_i=z_i)$. Therefore, the total computing resources consumed for local model training is given by
\begin{equation}\label{eq:computing}
\setlength{\abovedisplayskip}{1pt}
\setlength{\belowdisplayskip}{1pt}
C_{\text{total}}=\tau K\overline{C}_{\mathit{UE}},
\end{equation}
where $\tau$ and $K$ represent the total number of local trainings and communication rounds respectively. Armed with the above preparation, we are now starting to analyze how the resources affect the FL performance by evaluating local and global model accuracy.


\section{The Relationship between FL Performance and Consumed Resources}
Indeed, the unsuccessful transmission of local models in the uplink affects the aggregation of the global model, while the unsuccessful transmission in the downlink affects the updating and training of local models. Therefore, it is necessary to analyze how the computing resources and communication resources affect the FL performance by evaluating both the local and global model accuracy.

\subsection{Local Model Accuracy}
In an FL framework, no matter what local machine learning algorithm is used, each UE solves the local optimization problem for local training \cite{yang2020energy, LAC, HHY}, \emph{i.e.},
\begin{equation}\label{eq:LA}
\setlength{\abovedisplayskip}{1pt}
\setlength{\belowdisplayskip}{1pt}
\min_{h_i\in\mathbb{R}^d} G_i (w_r, h_i)\triangleq F_i(w_r+h_i)-(\nabla F_i(w_r)-\zeta\nabla F(w_r))^{\text{T}} h_i,
\end{equation}
where $\zeta$ is constant and $h_i$ represents the difference between the global model and the local model for UE $i$. Without loss of generality, we use the GD algorithm to update local models, as it can achieve the required high accuracy and facilitate the convergence analysis \cite{yang2020energy}, as follows,
\begin{equation}
\setlength{\abovedisplayskip}{1pt}
\setlength{\belowdisplayskip}{1pt}
h_i^{(r)(t+1)}=h_i^{(r)(t)}-\xi \nabla G_i(w_r, h_i^{(r)(t)}),
\end{equation}
where $\xi$ represents the step size and $h_i^{(r)(t)}$ denotes the value of $h_i$ at the $t$-th local iteration with given global model vector $w_r$. Moreover, $\nabla G_i(w_r, h_i^{(r)(t)})$ is the gradient of $G_i(w_r, h_i)$ at point $h_i=h_i^{(r)(t)}$. In addition, $w_i^r(t)=w_r+h_i^{(r)(t)}$ represents the local model of UE $i$ at the $t$-th local iteration. For a small step $\xi$, we can derive a set of solutions $h_i^{(r)(0)}, \cdots,h_i^{(r)(t)},\cdots,h_i^{(r)(\tau)}$, which satisfies 
\begin{equation}\notag
\setlength{\abovedisplayskip}{1pt}
\setlength{\belowdisplayskip}{1pt}
G_i(w_r, h_i^{(r)(0)})\geq \cdots \geq G_i(w_r, h_i^{(r)(t)})\geq \cdots \geq G_i(w_r, h_i^{(r)(\tau)}).
\end{equation}

To provide the convergence condition for the GD method, we introduce local model accuracy loss $\epsilon_l$ \cite{yang2020energy,LAC}, which resembles the approximate factors in [15] [24], as follows,
\begin{equation}\label{eq:epsilon_l}
\setlength{\abovedisplayskip}{1pt}
\setlength{\belowdisplayskip}{1pt}
G_i(w_r, h_i^{(r)(t)})-G_i(w_r, h_i^{(r)*})\leq
\epsilon_l(G_i(w_r, h_i^{(r)(0)})-G_i(w_r, h_i^{(r)*})), 
\end{equation}
where $h_i^{(r)*}$ represents the optimal solution of problem (\ref{eq:LA}). Note that each UE aims to solve the local optimization problem with a target local model accuracy $1-\epsilon_l$. To achieve the local model accuracy $1-\epsilon_l$ and the global model accuracy loss $1-\epsilon_g$ in the following, we first make the following three assumptions on the local loss function $F_i(w)$, as that in \cite{yang2020energy,Ajoint,HHY}.



$\bullet$~~Assumption 1: Function $F_i(w)$ is $L$-Lipschitz, \emph{i.e.}, $\forall w, w'\in \mathbb{R}^d, \parallel\nabla F_i(w)-\nabla F_i(w')\parallel\leq L\parallel w-w'\parallel.$

$\bullet$~~Assumption 2: Function $F_i(w)$ is $\gamma$-strongly convex, \emph{i.e.}, $\forall w, w'\in \mathbb{R}^d, F_i(w)\geq F_i(w')+\langle\nabla F_i(w'),(w-w')\rangle+ \frac{\gamma}{2}\parallel w-w'\parallel^2$.

$\bullet$~~Assumption 3: $F_i(w)$ is twice-continuously differentiable. And $\gamma I\leq \nabla^2 F_i(w)\leq LI$.

Based on the three assumptions, we can obtain the lower bound on the number of local iterations during each communication round, which is shown as Proposition 2.
\newtheorem{Proposition}{Proposition}
\begin{proposition} \label{Proposition1}
Local model accuracy loss $\epsilon_l$ is achieved if $\xi<\frac{2}{L}$ and run the GD method 
\begin{equation}\notag
\setlength{\abovedisplayskip}{2pt}
\setlength{\belowdisplayskip}{2pt}
\tau\geq \lceil \frac{2}{(2-L\xi)\xi\gamma}\ln \frac{1}{\epsilon_l}\rceil
\end{equation}
iterations during each communication round at each UE that participates in local training.
\end{proposition}
\begin{IEEEproof}
See Appendix B.
\end{IEEEproof}

The lower bound in Proposition 2 reflects the growing trend of the number of local iterations with respect to the local model accuracy, which can approximate the consumption of computing resources for training local models.

\vspace{-5mm}
\subsection{Global Model Accuracy}
In FL algorithms, a global model accuracy is also needed. For a specific FL task, we define $\epsilon_g$ as its global model accuracy loss (the global model accuracy is $1-\epsilon_g$), as follows,
\begin{equation}
\setlength{\abovedisplayskip}{2pt}
\setlength{\belowdisplayskip}{2pt}
 F(w_r(\hat{\cal S}, \mathit{SINR}_{\text{up}}))-F(w_r^*)\leq \epsilon_g ( F(w_1)-F(w_r^*)),
\end{equation}
where $w_r^*$ represents the actual optimal solution. Moreover,
we provide the following Proposition 3 about the number of communication rounds for achieving the global model accuracy $1-\epsilon_g$. 
\begin{proposition} \label{proposition 4}
Global model accuracy $1-\epsilon_g$ is achieved if the number of communication rounds $K$ meets 
\begin{equation}\notag
\setlength{\abovedisplayskip}{2pt}
\setlength{\belowdisplayskip}{2pt}
K\geq \lceil \frac{2L^2\ln \frac{1}{\epsilon_g}}{(1-\epsilon_l)\gamma^2\zeta}\rceil,
\end{equation}
when running FL algorithm shown as \text{Algorithm 1} with $0<\zeta<\frac{\gamma}{L}$

\end{proposition}
\begin{IEEEproof}
See Appendix C.
\end{IEEEproof}

Note that it is very hard to derive a closed-form expression of the global model during each communication round due to the dynamic nature of the wireless channel and the uncertain nature of multiple random variables. Therefore, we assume the amount of datasets on each UE is fixed to facilitate the proof of Proposition 3. In addition, from Proposition 2 and Proposition 3, we can see that there is a trade-off between the number of communication rounds and the number of local iterations characterized by $\epsilon_l$: small $\epsilon_l$ leads to large $\tau$, yet small $K$, from which we can jointly approximate the communication and computing resources consumed by training FL tasks. The details can be found in the next section.




\begin{algorithm}[!htbp]
    \caption{: FL Algorithm.}
    \label{FL_A}
    \hspace*{0.02in}{\bf Input:}
    The required local model accuracy loss $\epsilon_l$, the required global model accuracy loss $\epsilon_g$.\\
   \hspace*{0.02in}{\bf output:}
    The global model $w_r$, the number of local iterations $\tau$, and the number of communication rounds $K$.
    \begin{algorithmic}[1]
    \STATE
     Initialization: local models $w_i^1(0)=0$, the global model $g_1=0$.\\
    \FOR {$r=1,2,...$}
    \STATE Each UE calculates $\nabla F_i(w_r)$.
    \STATE Each UE sends $\nabla F_i(w_r)$ to the BS.
    \STATE The BS calculates $\nabla F(w_r)=\frac{\sum_{i=1}^n\hat{{\cal S}_i}\nabla F_i(w_r)}{\sum_{i=1}^n\hat{{\cal S}_i}}$.
    \STATE The BS broadcasts $\nabla F(w_r)$ to each UE.\\
    \STATE {\bf Parallel} Each UE $i=1,2,...,n$
    \STATE Initialization: the local iteration number $t=0$, and set $h_i^{(r)(0)}=0$.
    \STATE {\bf Repeat} 
    \STATE Every $V$ steps set $h_i^{(r)*}=h_i^{(r)(t)}$.
    \STATE Update $h_i^{(r)(t+1)}=h_i^{(r)(t)}-\xi \nabla G_i(w_r, h_i^{(r)(t)})$.
    \STATE Set $w_i^r(t)=w_r+h_i^{(r)(t)}$.
    \IF {$\frac{G_i(w_r, h_i^{(r)(t)})-G_i(w_r, h_i^{(r)*})}{(G_i(w_r, h_i^{(r)(0)})-G_i(w_r, h_i^{(r)*}))} > \epsilon_l$}
    \STATE Set $t=t+1$ 
    \ELSE
    \STATE  Each UE $i$ sends $w_i^r(t)$ to the BS.
    \ENDIF
    \STATE The BS calculates $w_r=\frac{\sum_{i=1}^n\hat{{\cal S}_i}w_i^r(t)}{\sum_{i=1}^n\hat{{\cal S}_i}}$
    \STATE The BS sends $w_r$ to all UEs.
    \IF {$\frac{F(w_r(\hat{\cal S}, \mathit{SINR}_{\text{up}}))-F(g^*)}{F(g_0)-F(g^*)}<\epsilon_g$}
    \STATE Break;
    \ENDIF
    \ENDFOR
    \STATE Set $\tau=t$, $K=r$.
   \STATE \textbf{output} $w_r$, $\tau$, $K$.
    \end{algorithmic}
    \end{algorithm}

\section{Discussions of Three cases}
In general, the resources used for training FL tasks at the wireless edge network should be limited, as 1) Communication and computing resources at the wireless edge network are limited and precious. 2) Resource consumption quickly increases with the widespread use of smart terminals. In this section, we discuss three specific cases for different sufficiency of communication and computing resources. Furthermore, we derive the explicit expression of the model accuracy under FL framework, as a function of the amount of the consumed resources based on the sufficiency of respective communication and computing resources.

\subsection{Case 1: Sufficient Communication Resources and Computing Resources}
When both communication resources and computing resources are sufficient, we can approximate the amount of communication/computing resources needed for the FL algorithm based on Proposition 2 and Proposition 3. Specifically, the bandwidth needed for transmitting local models in the uplink should meet
\begin{small}
\begin{equation}
\begin{split}
\setlength{\abovedisplayskip}{1pt}
\setlength{\belowdisplayskip}{1pt}
\overline{B}_{\text{up}}
\geq \lceil\frac{2L^2\ln \frac{1}{\epsilon_g}}{(1-\epsilon_l)\gamma^2\zeta}\rceil\cdot\sum_{i=1}^{n} b_{\text{up}}^{i,r} f_{\mathit{SINR}_{\text{up}}}~~~~~~~~~~~~~~~~~~~~~~~~~~~~~~~~~~~~~~~~~~~~~~~~~~~~~~~~~~~~~~~~~~~~~~~~~~~~~~~~~~~\\
=\lceil \frac{2L^2\ln \frac{1}{\epsilon_g}}{(1-\epsilon_l)\gamma^2\zeta}\rceil\cdot\sum_{i=1}^{n}\frac{s}{T_{\text{up}}^{i,r}\log_2(1+\frac{P_{\text{up}}G(d_1)}{\sum_{j=1}^{N_I}PG({d_2}^{(j)})+\delta^2})}\cdot
(\frac{2d_1}{(d_0)^2})\cdot\sum_{n_{\cal A}=n_I}^NC_{n_{\cal A}}^{n_I}(1-\text{exp}\{1-2t_{\text{up}}\lambda_a\})^{n_I}\cdot\\
(\text{exp}\{1-2t_{\text{up}}\lambda_a\})^{n_{\cal A}-n_I}\cdot
\frac{(\pi (d_0)^2 \lambda_i)^{n_{\cal A}}}{n_{\cal A}!}
\text{exp}\{-\pi (d_0)^2 \lambda_i\}\cdot
\left(\frac{2}{(d_0)^{2}}\right)^{n_I}\prod^{n_I}_{n=1} d_2^{(n)},
\end{split}
\end{equation} 
\end{small}

Similarly, based on equation (\ref{eq:BRD}), we can obtain the bandwidth needed for transmitting the global model in the downlink, as follows,
\begin{small}
\begin{equation}
\begin{split}
\setlength{\abovedisplayskip}{1pt}
\setlength{\belowdisplayskip}{1pt}
\overline{B}_{\text{down}}
\geq \lceil\frac{2L^2\ln \frac{1}{\epsilon_g}}{(1-\epsilon_l)\gamma^2\zeta}\rceil\cdot \sum_{i=1}^{n} b_{\text{down}}^{i,r} f_{\mathit{SNR}_{\text{down}}}
=\lceil\frac{2L^2\ln \frac{1}{\epsilon_g}}{(1-\epsilon_l)\gamma^2\zeta}\rceil\cdot\sum_{i=1}^{n}\frac{s}{T_{\text{down}}^{i,r}\log_2(1+\frac{P_{\text{down}}G(d_1)}{\delta^2})}\cdot\frac{2d_1}{r_0^2}.
\end{split}
\end{equation}
\end{small}

Furthermore, based on equation (19), given local accuracy $\epsilon_l$ with $\xi<\frac{2}{L}$, the total amount of computing resources should meet the following constraint,
\begin{small}
\begin{equation}
\begin{split}
\setlength{\abovedisplayskip}{2pt}
\setlength{\belowdisplayskip}{2pt}
C_{\text{total}} \geq 
\lceil \frac{2}{(2-L\xi)\xi\gamma}\ln \frac{1}{\epsilon_l}\rceil\cdot
\lceil \frac{2L^2\ln \frac{1}{\epsilon_g}}{(1-\epsilon_l)\gamma^2\zeta}\rceil\cdot
\sum_{i=1}^n\frac{c_iS_i}{T_i}\cdot\frac{1}{\sqrt{2\pi}\sigma_{i}}\text{exp}\left(-\frac{(s_i-\mu_{i})^2}{2\sigma_{i}^2}\right)\cdot\frac{(d_0)^2-(d_{\text{min}})^2}{(r_0)^2}.
\end{split}
\end{equation}
\end{small}

\vspace{-8mm}
\subsection{Case 2: Sufficient Computing Resources and Insufficient Communication Resources}
When computing resources are sufficient, while communication resources are insufficient, we aim to reduce bandwidth consumption by reducing the number of communication rounds. 
In this case, the number of local iterations still follows Proposition 2, as computing resources are sufficient. However, Proposition 3 may not be met due to the lack of communication resources, which decreases the number of communication rounds $K$. As a result, the required global model accuracy cannot be achieved. Specifically, the maximal number of communication rounds $K_{\text{max}}$ is limited by the communication resources, \emph{i.e.}, $K_{\text{max}}=\lfloor\min\{\frac{B_{\text{down}}^{\text{max}}}{\overline{B}_{\text{down}}^r},\frac{B_{\text{up}}^{\text{max}}}{\overline{B}_{\text{up}}^r}\}\rfloor$, where $B_{\text{up}}^{\text{max}}$ and $B_{\text{down}}^{\text{max}}$ are the maximal available bandwidth that can be used for FL in the uplink and the downlink respectively, $\overline{B}_{\text{up}}^r=\sum_{i=1}^{n} b_{\text{up}}^{i,r} f_{\mathit{SINR}_{\text{up}}}$ and $\overline{B}_{\text{down}}^r=\sum_{i=1}^{n} b_{\text{down}}^{i,r} f_{\mathit{SNR}_{\text{down}}}$ are the mean bandwidth consumption on the uplink and downlink at one global iteration, respectively. To achieve the required global accuracy $1-\epsilon_g$, when the number of communication round is limited, based on Appendix C, we first give the following relationship,
\begin{small}
\begin{equation}
\setlength{\abovedisplayskip}{1pt}
\setlength{\belowdisplayskip}{1pt}
F\left(g_{r}\left(\hat{\cal S}, \mathit{SINR}_{\text{up}}\right)\right)-F\left(g^*\right)
\leq \text{exp}\left(-K\left(\frac{(1-\epsilon_l)\gamma^2\zeta}{2L^2}\right)\right)\left(F(g_0)-F\left(g^*\right)\right),
\end{equation}
\end{small}
where we reasonably expect the real achieved global accuracy loss $\widetilde{\epsilon}_g$ can be expressed by
\begin{small}
\begin{equation}
\setlength{\abovedisplayskip}{1pt}
\setlength{\belowdisplayskip}{1pt}
\widetilde{\epsilon}_g=\text{exp}\left(-K\left(\frac{(1-\widetilde{\epsilon}_l)\gamma^2\zeta}{2L^2}\right)\right).
\end{equation}
\end{small}

Therefore, we have $K=\lceil\frac{2L^2\ln\frac{1}{\widetilde{\epsilon}_g}}{(1-\widetilde{\epsilon}_l)\gamma^2\zeta}\rceil$, 
where $\widetilde{\epsilon}_l$ is the realistic local model accuracy loss. Moreover, we can see that when $\widetilde{\epsilon}_g$ is fixed, $K$ will decrease if $\widetilde{\epsilon}_l$ decreases. If we want to reduce $K$ thus to reduce bandwidth consumption, while keeping $\widetilde{\epsilon}_g$ unchanged, we should decrease $\widetilde{\epsilon_l}$ by increasing the number of local iterations. As a result, the computing resource consumption will increase. In other words, there exists a trade-off to some extent between the communication resources and computing resources for achieving a certain ML model accuracy. In addition, from the perspective of communication resources, the number of communication rounds $K$ should meet $K\leq K_{\text{max}}$. Therefore, we have $\widetilde{\epsilon}_l\leq \lfloor1-\frac{2L^2\ln\frac{1}{\widetilde{\epsilon}_g}}{K_{\text{max}}\gamma^2\zeta}\rfloor$.

Therefore, based on Proposition 2, the number of local iterations should meet
\begin{small}
\begin{equation}
\setlength{\abovedisplayskip}{1pt}
\setlength{\belowdisplayskip}{1pt}
\tau \geq \lceil\frac{2}{(2-L\xi)\xi\gamma}\ln\frac{K_{\text{max}}\gamma^2\zeta}{K_{\text{max}}\gamma^2\zeta-2L^2\ln\frac{1}{\widetilde{\epsilon}_g}}\rceil.
\end{equation}
\end{small}

As a result, the total computing resources consumed for the FL task is given by
\begin{small}
\begin{equation}
\begin{split}
\setlength{\abovedisplayskip}{1pt}
\setlength{\belowdisplayskip}{1pt}
C_{\text{total}}\geq \sum_{i=1}^n\frac{c_iS_i}{T_i}\cdot\frac{1}{\sqrt{2\pi}\sigma_{i}}\text{exp}\left(-\frac{(s_i-\mu_{i})^2}{2\sigma_{i}^2}\right)\cdot\frac{(d_0)^2-(d_{\text{min}})^2}{(r_0)^2}\cdot\\ \lceil\frac{2}{(2-L\xi)\xi\gamma}\ln\frac{K_{\text{max}}\gamma^2\zeta}{K_{\text{max}}\gamma^2\zeta-2L^2\ln\frac{1}{\widetilde{\epsilon}_g}}\rceil\cdot
\lceil\frac{2L^2\ln\frac{1}{\widetilde{\epsilon}_g}}{(1-\widetilde{\epsilon}_l)\gamma^2\zeta}\rceil.
\end{split}
\end{equation}
\end{small}

\vspace{-8mm}
\subsection{Case 3: Sufficient Communication Resources and Insufficient Computing Resources}
When communication resources are sufficient, while computing resources are insufficient, we aim to reduce the computing resource consumption by reducing the number of local iterations. As communication resources are sufficient, the number of communication rounds still follows Proposition 3, while Proposition 2 may not be met due to the lack of computing resources, which decreases the number of local iterations. As a result, the required local model accuracy cannot be achieved. In addition, from the perspective of computing resources, the number of local iterations should meet $\tau\cdot \overline{C}_{\text{UE}}\leq \sum_{i=1}^n C_i$, where $C_i$ represents the maximal computing resources used for local training on UE $i$. To achieve the required local accuracy although the number of local iterations are limited, we give the following relationship based on Appendix B,
\begin{small}
\begin{equation}
\setlength{\abovedisplayskip}{0pt}
\setlength{\belowdisplayskip}{1pt}
 G_i\left(w_r, h_i^{(r)(t)}\right)-G_i\left(w_r, h_i^{(r)^*}\right)
 \leq \text{exp}\left(-\tau\frac{\left(2-L\xi\right)\xi\gamma}{2}\right)\cdot
 \left(G_i\left(w_r, h_i^{(r)(0)}\right)-G_i\left(w_r, h_i^{(r)^*}\right) \right),
\end{equation}
\end{small}
from which we can reasonably expect that the real local model accuracy loss $\widetilde{\epsilon}_l$ is expressed by $\widetilde{\epsilon}_l=\text{exp}\left(-\tau\frac{\left(2-L\xi\right)\xi\gamma}{2}\right)$, 
where we can obtain the number of local iterations, \emph{i.e.}, $\tau=\lceil\frac{2\ln\frac{1}{\widetilde{\epsilon}_l}}{(2-L\xi)\xi\gamma}\rceil$.


Therefore, when $\xi<\frac{2}{L}$, we have $\widetilde{\epsilon}_l \geq \text{exp}\left(\frac{(L\xi-2)\xi\gamma\sum_{i=1}^nC_i}{2\overline{C}_{\text{UE}}}\right)$.
In other words, when the total amount of available computing resource decreases, the lower bound of $\widetilde{\epsilon}_l$ will increase. Moreover, based on $\text{C}_{\text{UE}}$ in Section IV. D, we can derive the lower bound of the number of communication rounds, as follows,
\begin{small}
\begin{equation} \label{eq:Kwithtau}
\setlength{\abovedisplayskip}{1pt}
\setlength{\belowdisplayskip}{1pt}
K\geq \lceil \frac{2L^2\ln \frac{1}{\epsilon_g}}{\left(1-\text{exp}\left(\frac{(L\xi-2)\xi\gamma\sum_{i=1}^nC_i}{2\sum_{i=1}^n\frac{c_iS_i}{T_i}\cdot\frac{1}{\sqrt{2\pi}\sigma_{i}}\text{exp}\left(-\frac{(s_i-\mu_{i})^2}{2\sigma_{i}^2}\right)}\right)\right)\gamma^2\zeta}\rceil.
\end{equation}
\end{small}

Therefore, the bandwidth for transmitting local models and the global model in the uplink and the downlink are respectively given by
\begin{small}
\begin{equation}
\begin{split}
\setlength{\abovedisplayskip}{1pt}
\setlength{\belowdisplayskip}{1pt}
\overline{B}_{\text{up}}\geq \lceil \frac{2L^2\ln \frac{1}{\epsilon_g}}{\left(1-\text{exp}\left(\frac{(L\xi-2)\xi\gamma\sum_{i=1}^nC_i}{2\sum_{i=1}^n\frac{c_iS_i}{T_i}\cdot\frac{1}{\sqrt{2\pi}\sigma_{i}}\text{exp}\left(-\frac{(s_i-\mu_{i})^2}{2\sigma_{i}^2}\right)}\right)\right)\gamma^2\zeta}\rceil \cdot  \sum_{i=1}^{n}\frac{s}{T_{\text{up}}^{i,r}\log_2(1+\frac{P_{\text{up}}G(d_1)}{\sum_{j=1}^{N_I}PG({d_2}^{(j)})+\delta^2})}\cdot\\
(\frac{2d_1}{(d_0)^2})\cdot\sum_{n_{\cal A}=n_I}^NC_{n_{\cal A}}^{n_I}(1-\text{exp}\{1-2t_{\text{up}}\lambda_a\})^{n_I}\cdot
(\text{exp}\{1-2t_{\text{up}}\lambda_a\})^{n_{\cal A}-n_I}\cdot\\
\frac{(\pi (d_0)^2 \lambda_i)^{n_{\cal A}}}{n_{\cal A}!}
\text{exp}\{-\pi (d_0)^2 \lambda_i\}\cdot
\left(\frac{2}{(d_0)^{2}}\right)^{n_I}\prod^{n_I}_{n=1} d_2^{(n)},
\end{split}
\end{equation}
\end{small}
\begin{small}
\begin{equation}
\setlength{\abovedisplayskip}{1pt}
\setlength{\belowdisplayskip}{1pt}
\overline{B}_{\text{down}}\geq \lceil \frac{2L^2\ln \frac{1}{\epsilon_g}}{\left(1-\text{exp}\left(\frac{(L\xi-2)\xi\gamma\sum_{i=1}^nC_i}{2\sum_{i=1}^n\frac{c_iS_i}{T_i}\cdot\frac{1}{\sqrt{2\pi}\sigma_{i}}\text{exp}\left(-\frac{(s_i-\mu_{i})^2}{2\sigma_{i}^2}\right)}\right)\right)\gamma^2\zeta}\rceil \cdot \sum_{i=1}^{n}\frac{s}{T_{\text{down}}^{i,r}\log_2(1+\frac{P_{\text{down}}G(d_1)}{\delta^2})}\cdot\frac{2d_1}{(r_0)^2}.
\end{equation}
\end{small}

Therefore, based on the analysis aforementioned, we provide Proposition 4 about the resource consumption for the three cases discussed above. 
\begin{small}
\begin{proposition} \label{proposition 4}
\begin{enumerate}
\def\labelenumi{(\arabic{enumi})}
\item
Case 1-Sufficient Communication and Computing Resources: To achieve the required model accuracy $\epsilon_g$ and $\epsilon_l$, the consumption of bandwidth in the uplink is $\overline{B}_{\text{up}}
\geq \lceil \frac{2L^2\ln \frac{1}{\epsilon_g}}{(1-\epsilon_l)\gamma^2\zeta}\rceil\cdot\left(\frac{2}{(d_0)^{2}}\right)^{n_I}\prod^{n_I}_{n=1} d_2^{(n)}\cdot\sum_{i=1}^{n}\frac{s}{T_{\text{up}}^{i,r}\log_2(1+\frac{P_{\text{up}}G(d_1)}{\sum_{j=1}^{N_I}PG({d_2}^{(j)})+\delta^2})}\cdot(\frac{2d_1}{(d_0)^2})\cdot\sum_{n_{\cal A}=n_I}^NC_{n_{\cal A}}^{n_I}(1-\text{exp}\{1-2t_{\text{up}}\lambda_a\})^{n_I}\cdot(\text{exp}\{1-2t_{\text{up}}\lambda_a\})^{n_{\cal A}-n_I}\cdot\frac{(\pi (d_0)^2 \lambda_i)^{n_{\cal A}}}{n_{\cal A}!}\text{exp}\{-\pi (d_0)^2 \lambda_i\}$,
the consumption of bandwidth in the downlink is $\overline{B}_{\text{down}}
\geq \lceil\frac{2L^2\ln \frac{1}{\epsilon_g}}{(1-\epsilon_l)\gamma^2\zeta}\rceil\cdot\frac{2d_1}{r_0^2}\cdot\sum_{i=1}^{n}\frac{s}{T_{\text{down}}^{i,r}\log_2(1+\frac{P_{\text{down}}G(d_1)}{\delta^2})}$,
and the consumption of computing resources used for local training is $C_{\text{total}} \geq 
\lceil \frac{2}{(2-L\xi)\xi\gamma}\ln \frac{1}{\epsilon_l}\rceil\cdot
\lceil \frac{2L^2\ln \frac{1}{\epsilon_g}}{(1-\epsilon_l)\gamma^2\zeta}\rceil\cdot
\sum_{i=1}^n\frac{c_iS_i}{T_i}\cdot\frac{1}{\sqrt{2\pi}\sigma_{i}}\text{exp}\left(-\frac{(s_i-\mu_{i})^2}{2\sigma_{i}^2}\right)\cdot\frac{(d_0)^2-(d_{\text{min}})^2}{(r_0)^2}.$

\item
Case 2-Sufficient Computing Resources and Insufficient Communication Resources: To achieve the required global model accuracy $\epsilon_g$, the consumption of computing resources is $C_{\text{total}}\geq \sum_{i=1}^n\frac{c_iS_i}{T_i}\cdot\frac{1}{\sqrt{2\pi}\sigma_{i}}\text{exp}\left(-\frac{(s_i-\mu_{i})^2}{2\sigma_{i}^2}\right)\cdot\frac{(d_0)^2-(d_{\text{min}})^2}{(r_0)^2}\cdot \lceil\frac{2}{(2-L\xi)\xi\gamma}\ln\frac{K_{\text{max}}\gamma^2\zeta}{K_{\text{max}}\gamma^2\zeta-2L^2\ln\frac{1}{\widetilde{\epsilon}_g}}\rceil\cdot
\lceil\frac{2L^2\ln\frac{1}{\widetilde{\epsilon}_g}}{(1-\widetilde{\epsilon}_l)\gamma^2\zeta}\rceil.$

\item
Case 3-Sufficient Communication Resources and Insufficient Computing Resources: To achieve the required global model accuracy $\epsilon_g$, the consumption of bandwidth in the uplink is \\
$\overline{B}_{\text{up}}\geq \lceil \frac{2L^2\ln \frac{1}{\epsilon_g}}{\left(1-\text{exp}\left(\frac{(L\xi-2)\xi\gamma\sum_{i=1}^nC_i}{2\sum_{i=1}^n\frac{c_iS_i}{T_i}\cdot\frac{1}{\sqrt{2\pi}\sigma_{i}}\text{exp}\left(-\frac{(s_i-\mu_{i})^2}{2\sigma_{i}^2}\right)}\right)\right)\gamma^2\zeta}\rceil \cdot  \sum_{i=1}^{n}\frac{s}{T_{\text{up}}^{i,r}\log_2(1+\frac{P_{\text{up}}G(d_1)}{\sum_{j=1}^{N_I}PG({d_2}^{(j)})+\delta^2})}\cdot\\
(\frac{2d_1}{(d_0)^2})\cdot\sum_{n_{\cal A}=n_I}^NC_{n_{\cal A}}^{n_I}(1-\text{exp}\{1-2t_{\text{up}}\lambda_a\})^{n_I}\cdot
(\text{exp}\{1-2t_{\text{up}}\lambda_a\})^{n_{\cal A}-n_I}\cdot
\frac{(\pi (d_0)^2 \lambda_i)^{n_{\cal A}}}{n_{\cal A}!}
\text{exp}\{-\pi (d_0)^2 \lambda_i\}\cdot
\left(\frac{2}{(d_0)^{2}}\right)^{n_I}\prod^{n_I}_{n=1} d_2^{(n)}$, while the consumption of bandwidth in the downlink is\\ $\overline{B}_{\text{down}}\geq \lceil \frac{2L^2\ln \frac{1}{\epsilon_g}}{\left(1-\text{exp}\left(\frac{(L\xi-2)\xi\gamma\sum_{i=1}^nC_i}{2\sum_{i=1}^n\frac{c_iS_i}{T_i}\cdot\frac{1}{\sqrt{2\pi}\sigma_{i}}\text{exp}\left(-\frac{(s_i-\mu_{i})^2}{2\sigma_{i}^2}\right)}\right)\right)\gamma^2\zeta}\rceil \cdot \sum_{i=1}^{n}\frac{s}{T_{\text{down}}^{i,r}\log_2(1+\frac{P_{\text{down}}G(d_1)}{\delta^2})}\cdot\frac{2d_1}{(r_0)^2}$. 

\end{enumerate}
\end{proposition}
\end{small}

\section{Numerical Results and Discussion}
In this section, we verify our analytical modeling using numerical simulations by (1) verifying the analytical results of transmission success probability (uplink and downlink) and resource (bandwidth resource and computing resource) consumption; (2) Measuring the performance of FL settings; (3) Examining the trade-off between the computing resources and communication resources under FL framework.

\subsection{Simulation Setting}
We consider an FL empowered wireless network composed of multiple UEs that are randomly generated and one central BS with a cloud server that serves as the FL model aggregator. The coverage of the BS is a circular area with a radius of $1\text{KM}$. The radius of the interfering area is set to $100$m. The transmit power of UEs and the serving BS is set to $20$dBm and $43$dBm respectively \cite{SunBC}. Moreover, the noise power is set to $-173$dBm \cite{SunBC}. The density of interfering UEs $\lambda_a$ is set to $1~\text{UE}/\text{m}^2$ and $t_{up}$ is randomly chosen within $[1,3]$ms. The path loss is modeled as $g(D_1)=34+40\log(D_1)$ \cite{liu2020device}. The number of CPU cycles required for computing one sample data is randomly chosen within $[1,4]\times10^4$~cycles/sample \cite{yang2020energy}. $\mu_i$ and $\sigma_i$ are randomly chosen within $[1000, 10000]$ and $[0.2,0.5]$. 

We consider using FL to solve the multi-class classification problem over MNIST datasets \cite{lecun1998mnist} for model training, where all datasets of UEs are randomly splitted with 75-25 ratio, for training and testing respectively \cite{pmlr-v54-mcmahan17a}. Moreover, we use a fully-connected two-layer network built over PyTorch, where the size of input layer, hidden layer and output layer is set to 784 ($28\times28$), 40 and 10 respectively. The activation function is ReLU, as it can greatly accelerate the convergence of gradient descent and increase the number of the activated neurons \cite{li2017convergence,krizhevsky2012imagenet}. In addition, a constant learning rate has always been the conventional choice \cite{darken1990note,liu2019variance,liu2021access}. Inspired by the hyper-parameter analysis and the corresponding experimental results in \cite{darken1990note,liu2019variance}, we set learning rate $\xi=0.03$. In addition, according to our neural network settings, the transmitted model size $s$ is around 1.156 MB, when using 32-bit floating-point computing. In addition, we set $L=1/10$, $\gamma=1/10$, $\xi=1/10$, $\zeta=1/10$ \cite{yang2020energy}.



\subsection{Simulation Results}
\subsubsection{Verifying Analytical Results}
First, we examine the local and the global model transmission success probability with varying UE density respectively. In the two simulations, based on PPP model, we randomly generated 30 specific point distributions for each UE density (0.1 intervals), where both the simulation results of ${\rm Pr}(\text{SINR}_\text{up}>\beta_\text{up})$ and ${\rm Pr}(\text{SNR}_\text{down}>\beta_\text{down})$ are averaged over these 30 different channel instances for each UE density. Fig. 3 and Fig. 4 show the probability ${\rm Pr}(\text{SINR}_\text{up}>\beta_\text{up})$ in the uplink and ${\rm Pr}(\text{SNR}_\text{down}>\beta_\text{down})$ in the downlink respectively for both analytical and simulation results with varying UE density under different threshold parameters. The analytical results of ${\rm Pr}(\text{SINR}_\text{up}>\beta_\text{up})$ and ${\rm Pr}(\text{SNR}_\text{down}>\beta_\text{down})$ are computed based on equation (14) and equation (15) respectively. From Fig.3 and Fig. 4, we can see that the curves of analytical results match closely to simulations for both the uplink and the downlink. Moreover, we can see that the smaller threshold ($\beta_\text{up}$ in Fig. 3, $\beta_\text{down}$ in Fig. 4), the larger transmission success probability.

\begin{figure*}[htbp]
\vspace*{-8mm}
  \centering
  \begin{minipage}{0.32\linewidth}
    \includegraphics[width=0.9\linewidth]{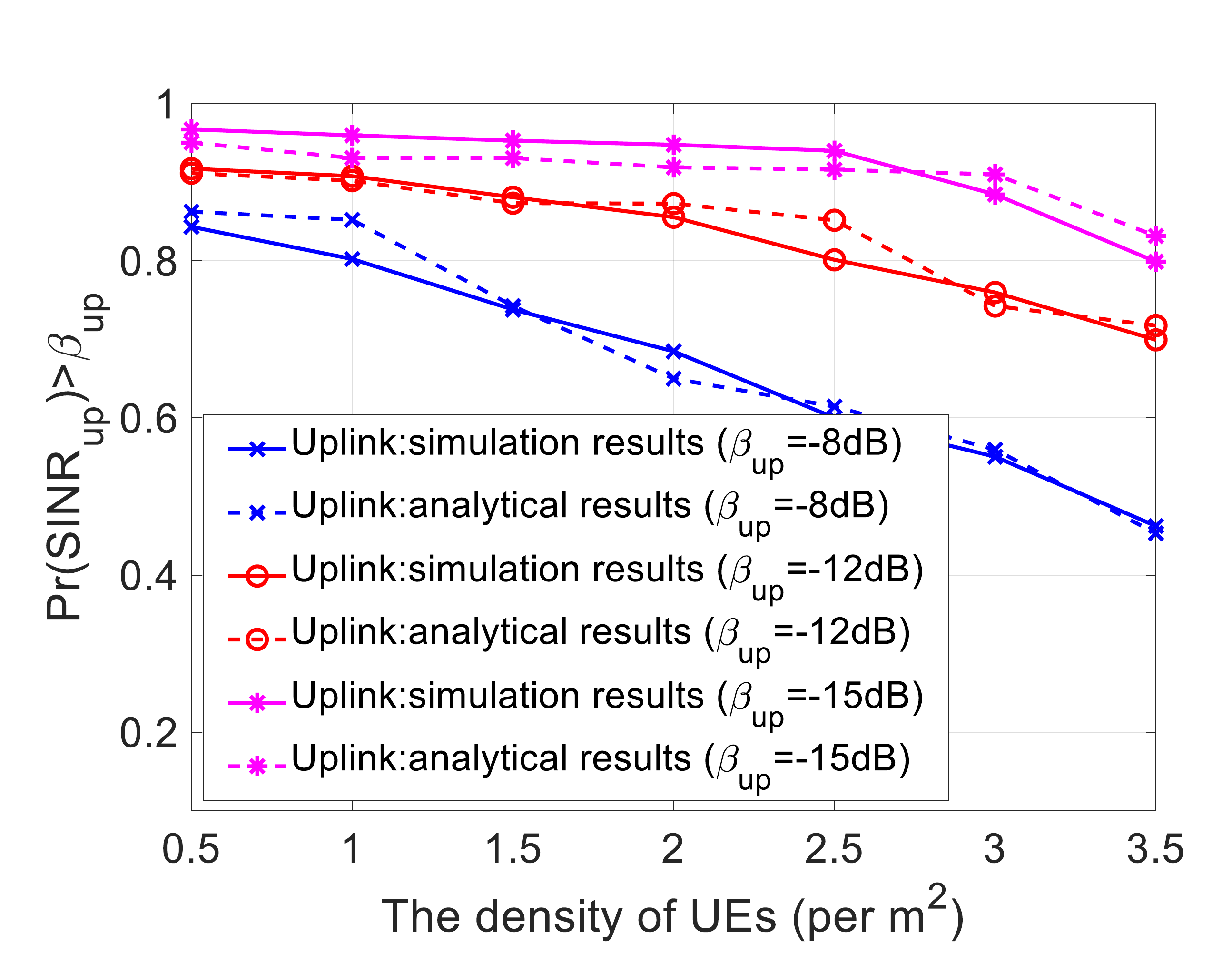}
    \caption{Comparison of the probability of successful transmission in the uplink.}
    \label{Fig4}
  \end{minipage}
  \begin{minipage}{0.32\linewidth}
    \centering
    \includegraphics[width=0.9\linewidth]{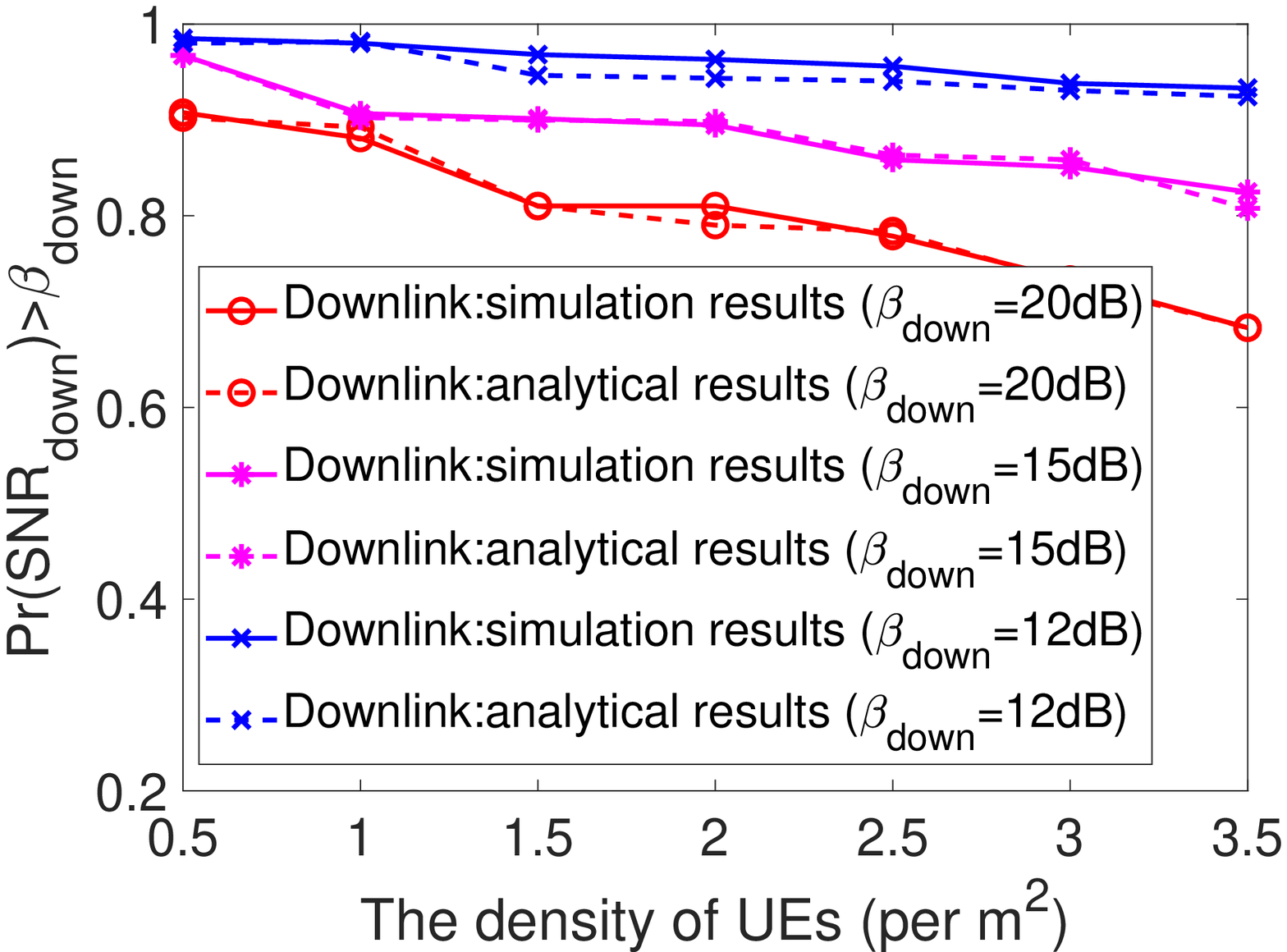}
    \caption{Comparison of the probability of successful transmission in the downlink.}
    \label{Fig5}
  \end{minipage}
  \begin{minipage}{0.32\linewidth}
    \centering
    \includegraphics[width=0.9\linewidth]{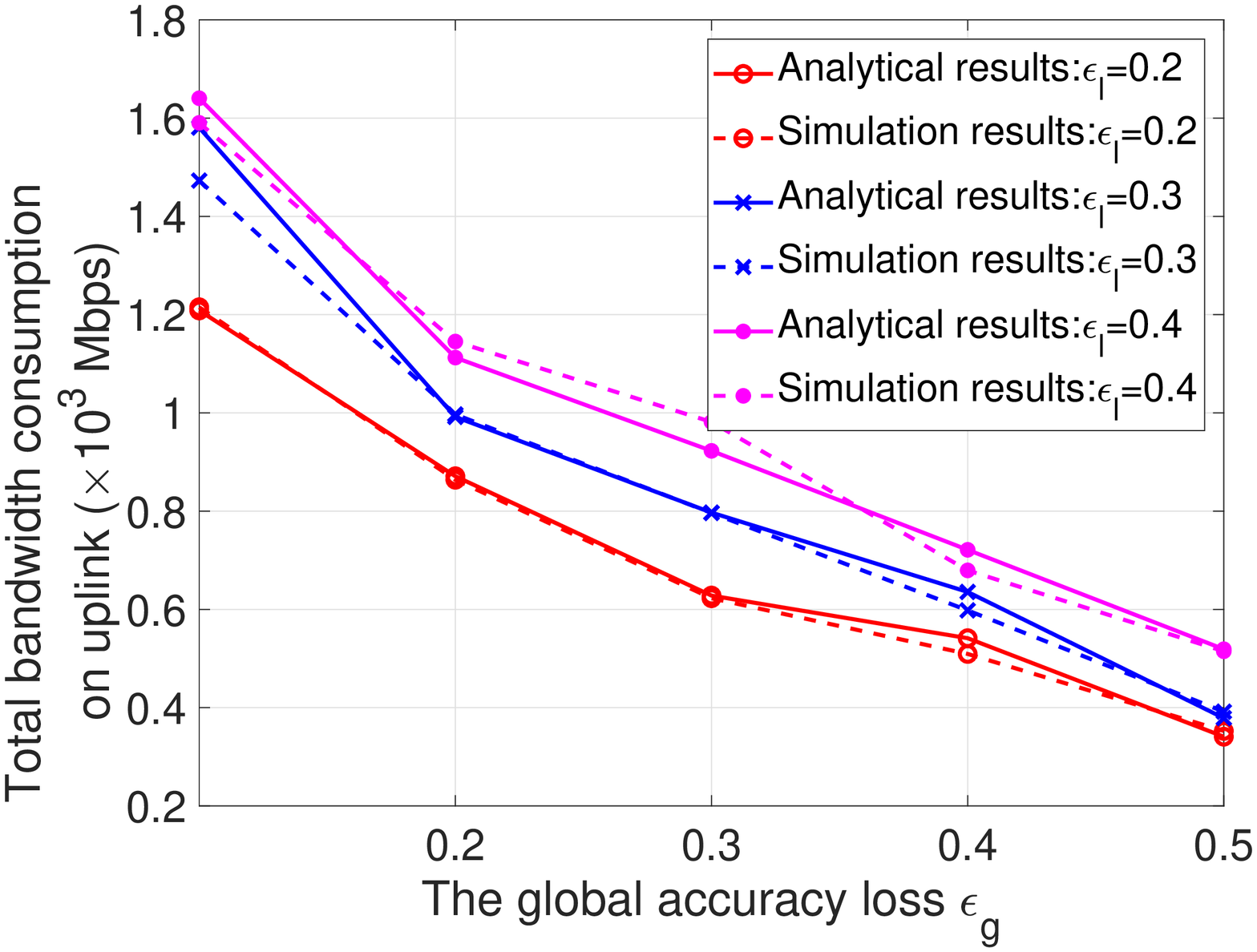}
    \caption{Comparison of bandwidth consumption in the uplink.}
    \label{UL}
  \end{minipage}
  \vspace*{-8mm}
\end{figure*}




Next, we examine the bandwidth consumption in the uplink and the downlink respectively for both analytical and simulation results with respect to the global accuracy loss $\epsilon_g$. We first randomly select UE density within $[1, 2]$ and randomly select 10 specific point distributions under the corresponding UE density. Then, we train the same FL task (\emph{i.e.}, classification on MNIST datasets) for each point distribution, where the simulation results are averaged over 10 point distributions. Fig. 5 and Fig. 6 show the bandwidth consumption in the uplink and the downlink changes with the global accuracy loss respectively. From Fig. 5 and Fig. 6, we can see that the curves of analytical results match closely to simulations for both the uplink and the downlink. Moreover, both the bandwidth consumption in the uplink and the downlink decrease with respect to the global accuracy loss. In addition, we also find that the lower local accuracy leads to more bandwidth consumption to guarantee a specific global accuracy when training i.i.d data. Specifically, as shown in Fig. 5, in the uplink, the amount of bandwidth consumed to guarantee $\epsilon_l=0.3$ is $0.144\times10^3$ Mbps more than that to guarantee $\epsilon_l=0.2$ on average, while the amount of bandwidth consumed to guarantee $\epsilon_l=0.4$ is $0.089\times10^3$ Mbps more than that to guarantee $\epsilon_l=0.3$ on average. As shown in Fig. 6, in the downlink, the amount of bandwidth consumed to guarantee $\epsilon_l=0.3$ is $0.54\times 10^4$ Mbps more than that to guarantee $\epsilon_l=0.2$ on average, while the amount of bandwidth to guarantee $\epsilon_l=0.4$ is $0.28\times 10^4$ Mbps more than that to guarantee $\epsilon_l=0.3$ on average. The reason is that the lower local accuracy needs more communication rounds to aggregate the local models to achieve a certain global accuracy, and thus consumes more bandwidth.

\begin{figure*}[htbp]
\vspace*{-6mm}
  \centering
  \begin{minipage}{0.32\linewidth}
    \includegraphics[width=0.9\linewidth]{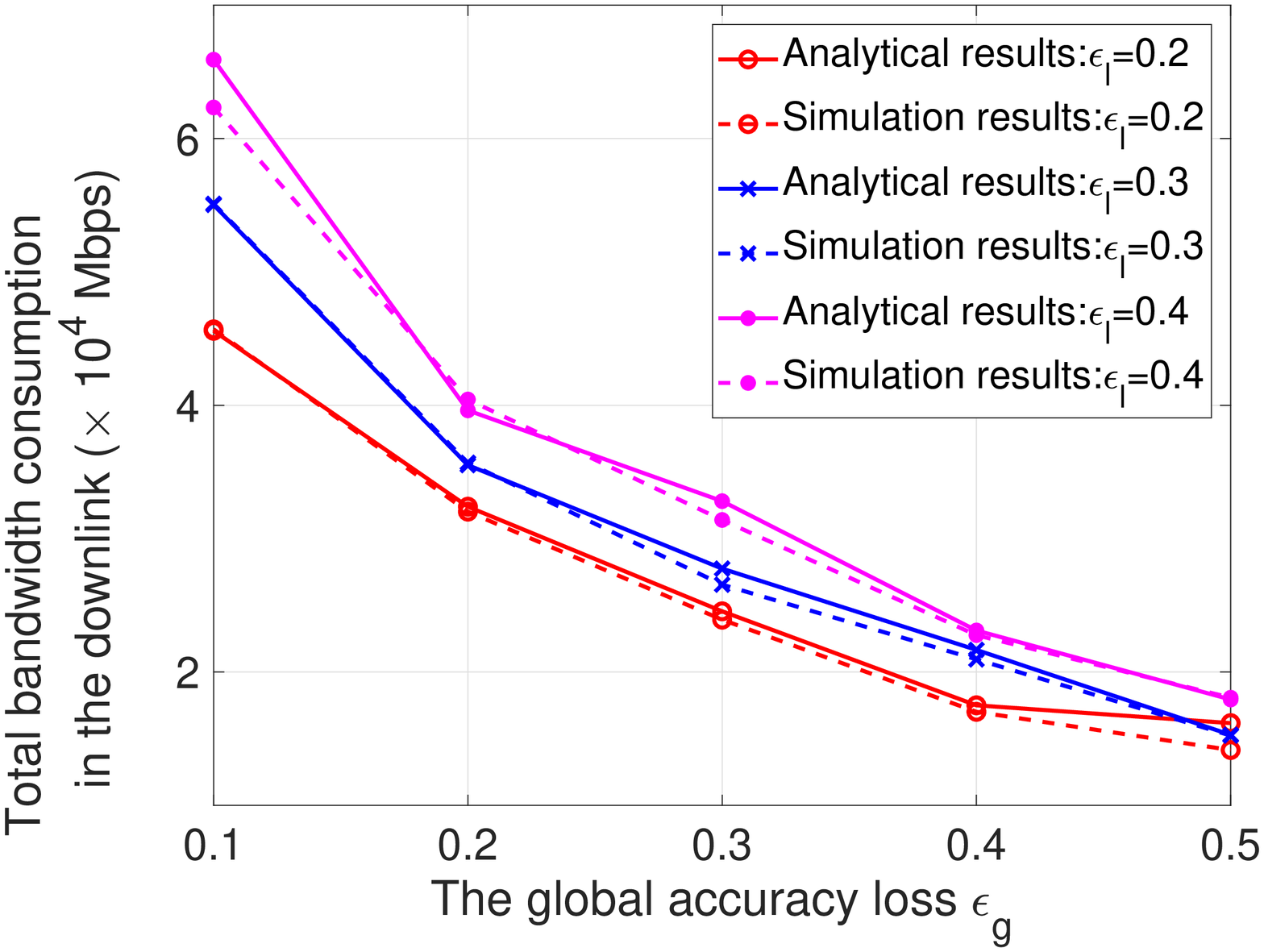}
    \caption{Comparison of bandwidth consumption in the downlink.}
    \label{DL}
  \end{minipage}
  \begin{minipage}{0.32\linewidth}
    \centering
    \vspace*{-2mm}
    \includegraphics[width=0.9\linewidth]{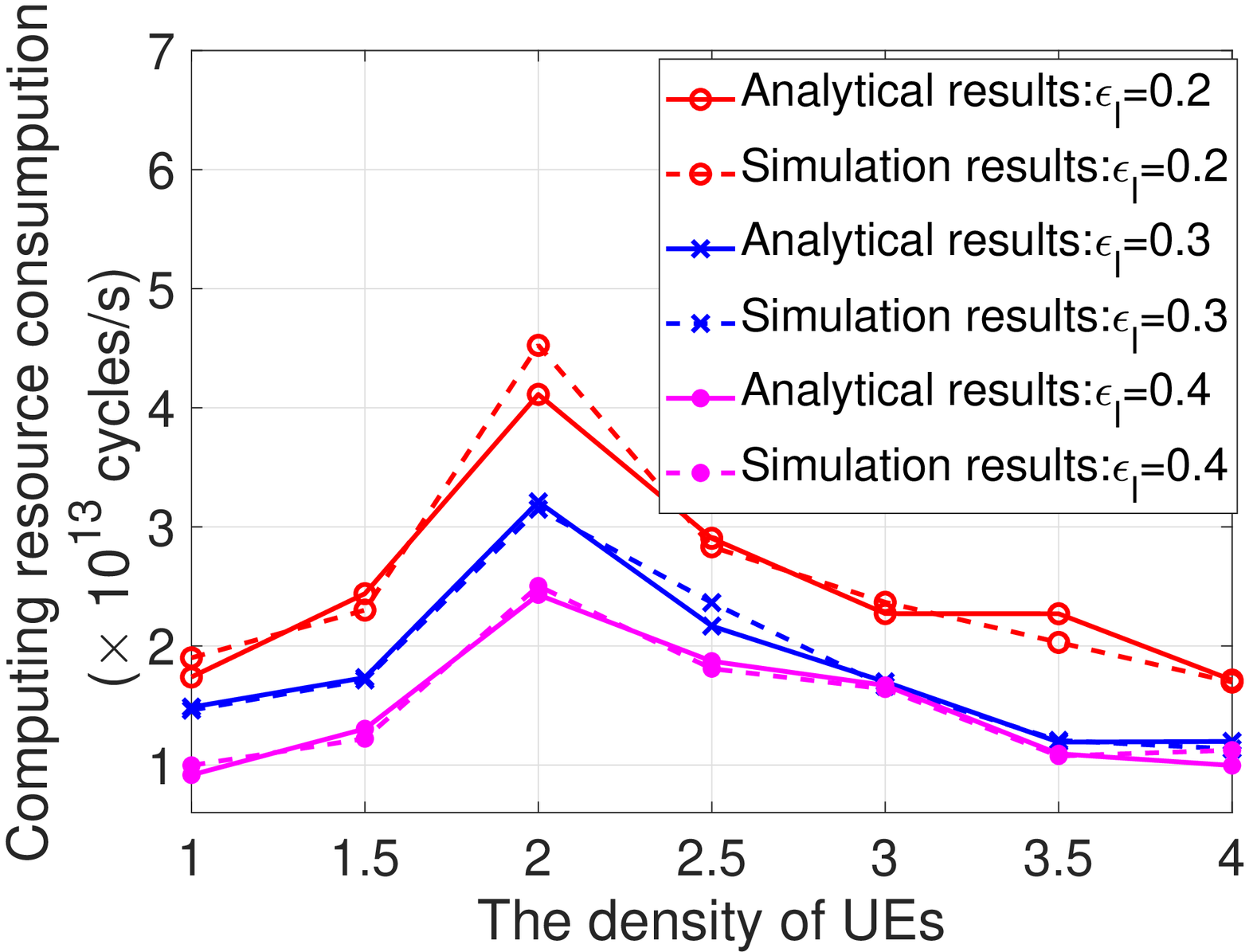}
    \caption{Comparison of the computing resource consumption.}
    \label{Compare}
  \end{minipage}
    \begin{minipage}{0.32\linewidth}
    \centering
    \includegraphics[width=0.9\linewidth]{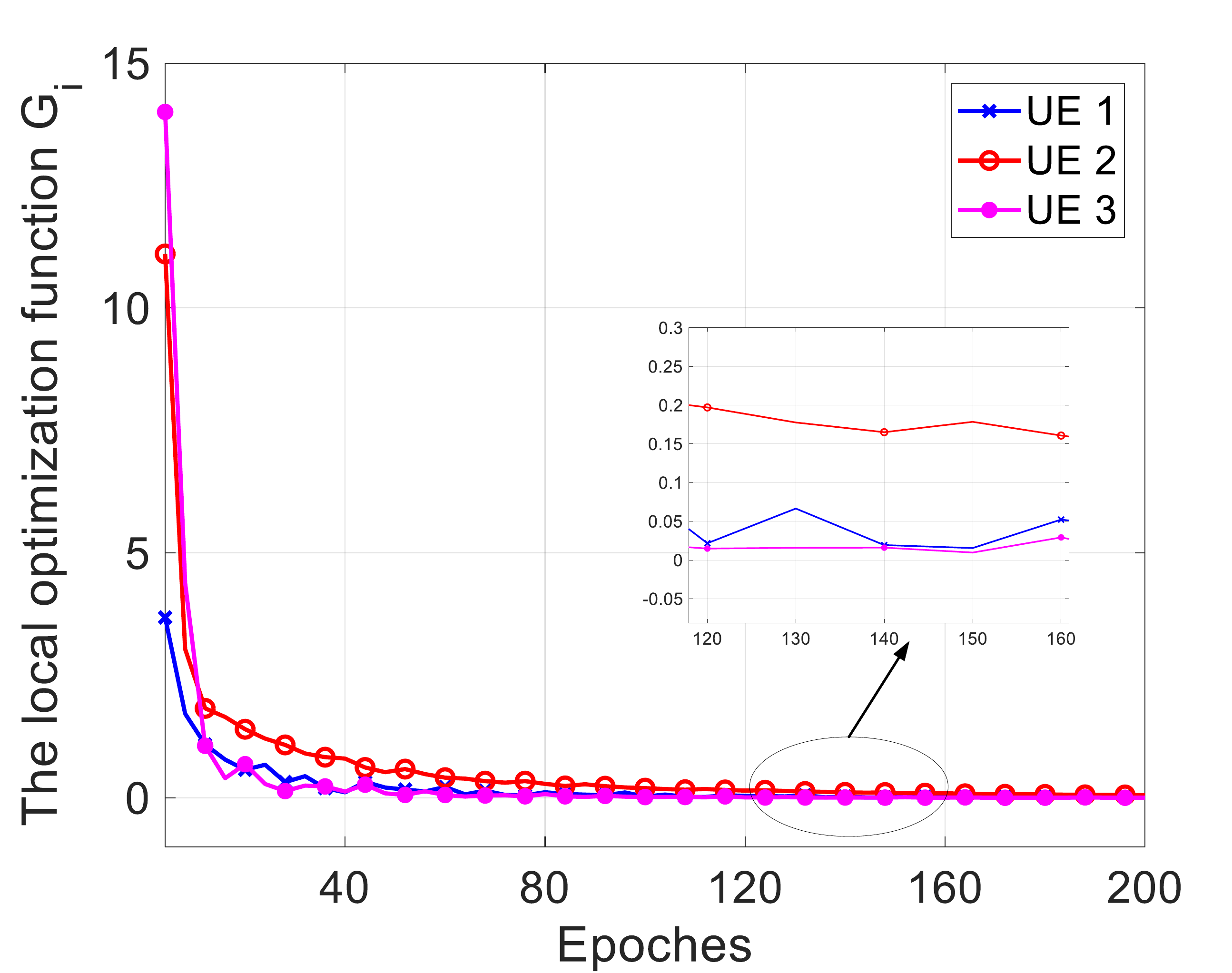}
    \caption{Local training during each communication round.}
    \label{Compare}
  \end{minipage}
\vspace*{-5mm}
\end{figure*}




 In the following, we examine the computing resource consumption for both analytical and simulation results with respect to the density of UEs. Specifically, the analytical results of computing resource consumption is computed based on equation (19), while the simulation results are averaged over 10 randomly generated point distributions for each UE density. Fig. 7 shows the computing resource consumption changes with the density of UEs. From Fig. 7, we can see that the amount of computing resource consumption increases in the beginning and then decreases with the density of UEs. Specifically, the amount of computing resource consumption increases with UE density, when it is approximately below 2 (\textit{i.e.}, $\lambda_i\leq 2$), as the number of UEs that participate in local training increases with UE density. When approximately $\lambda_i\geq 2$, the amount of computing resource consumption decreases with UE density, as poor SNR causes that some UEs fail in successfully receiving the global model. As a result, the number of UEs that participate in local training decreases in the next communication round, and thus the amount of computing resource consumption decreases. Moreover, we also find that achieving higher local accuracy needs more computing resources to train local models. Specifically, the amount of computing resources consumed to guarantee $\epsilon_l=0.2$ is $0.27\times 10^{12}$ cycles/s more than that to guarantee $\epsilon_l=0.3$ on average, while the amount of computing resources consumed to guarantee $\epsilon_l=0.3$ is $15\times 10^{12}$ cycles/s more than that to guarantee $\epsilon_l=0.4$ on average.

\subsubsection{Measuring the performance of FL settings}
First, we examine the convergence property by using simulation experiments. In this simulation experiment, the UE density is randomly chosen within $[1, 2]$ and data points are randomly generated (the same settings for the following simulations). Moreover, we set $\epsilon_g=0.2$, $\beta_\text{up}=-15$dB, and $\beta_\text{down}=15$dB. As shown in Fig. 8, we randomly choose $3$ UEs to observe the changes of the local optimization function, where the local optimization function converges in about 40 epochs. In addition, as shown in Fig. 9, we can observe that the global loss function convergences in around 12 communication rounds.


\begin{figure*}[htbp]
\vspace*{-5mm}
  \centering
  \begin{minipage}{0.32\linewidth}
    \centering
    \vspace{-5mm}
    \includegraphics[width=0.9\linewidth]{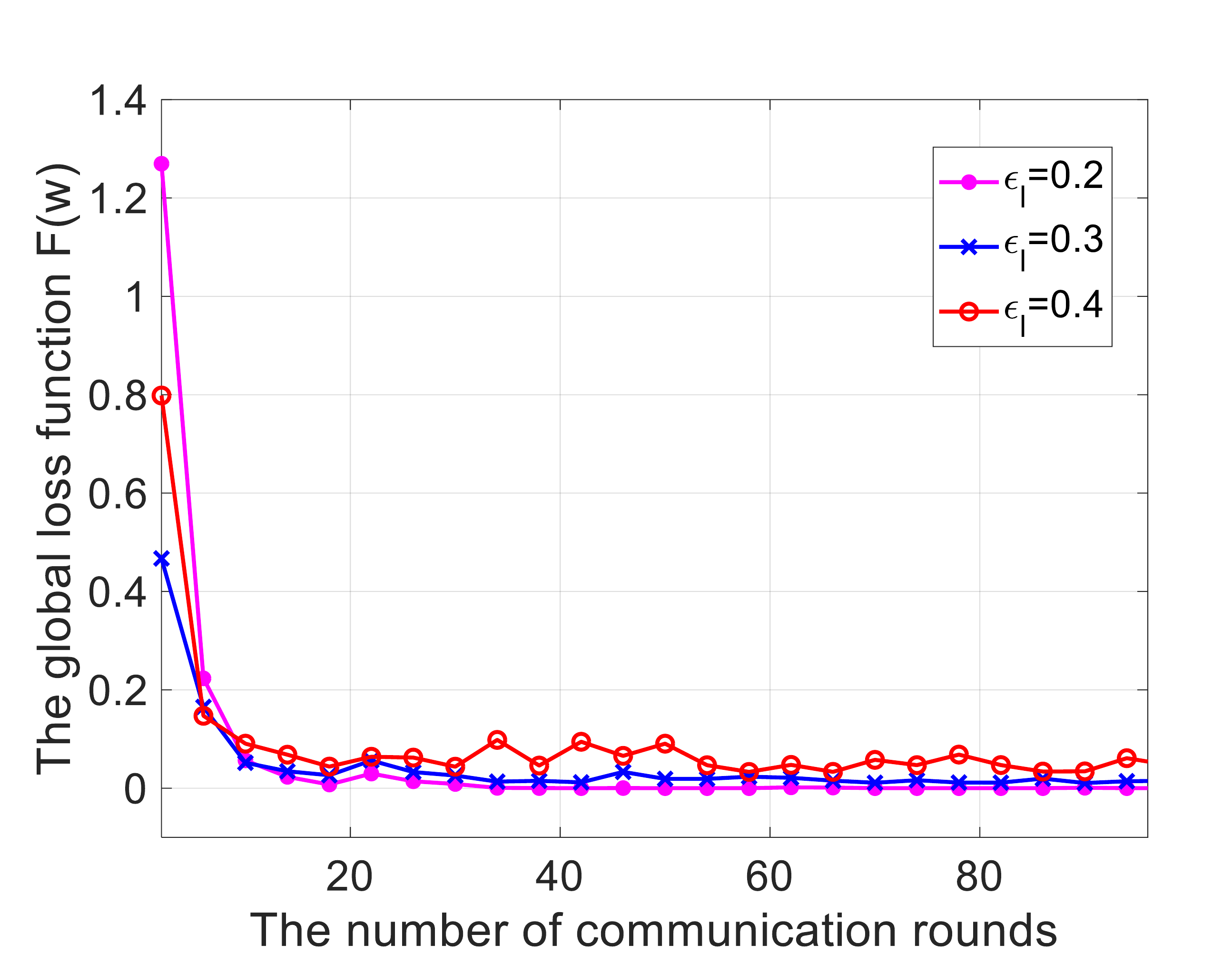}
    \caption{Convergence.}
    \label{UL}
  \end{minipage}
  \begin{minipage}{0.32\linewidth}
    \centering
    \vspace{-5mm}
    \includegraphics[width=0.9\linewidth]{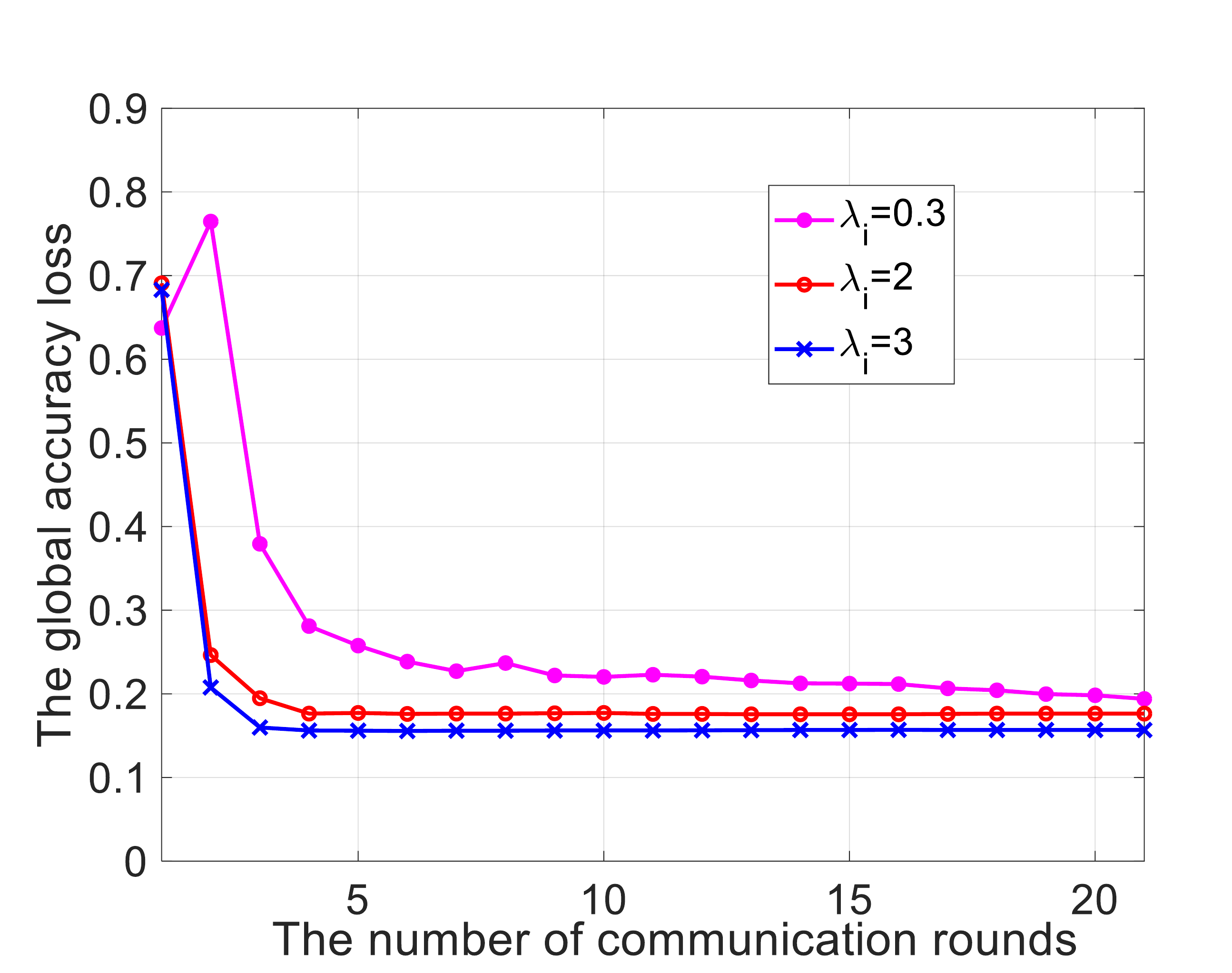}
    \caption{Comparison of the global accuracy loss.}
    \label{fig10}
  \end{minipage}
  \begin{minipage}{0.32\linewidth}
    \centering
    \includegraphics[width=0.9\linewidth]{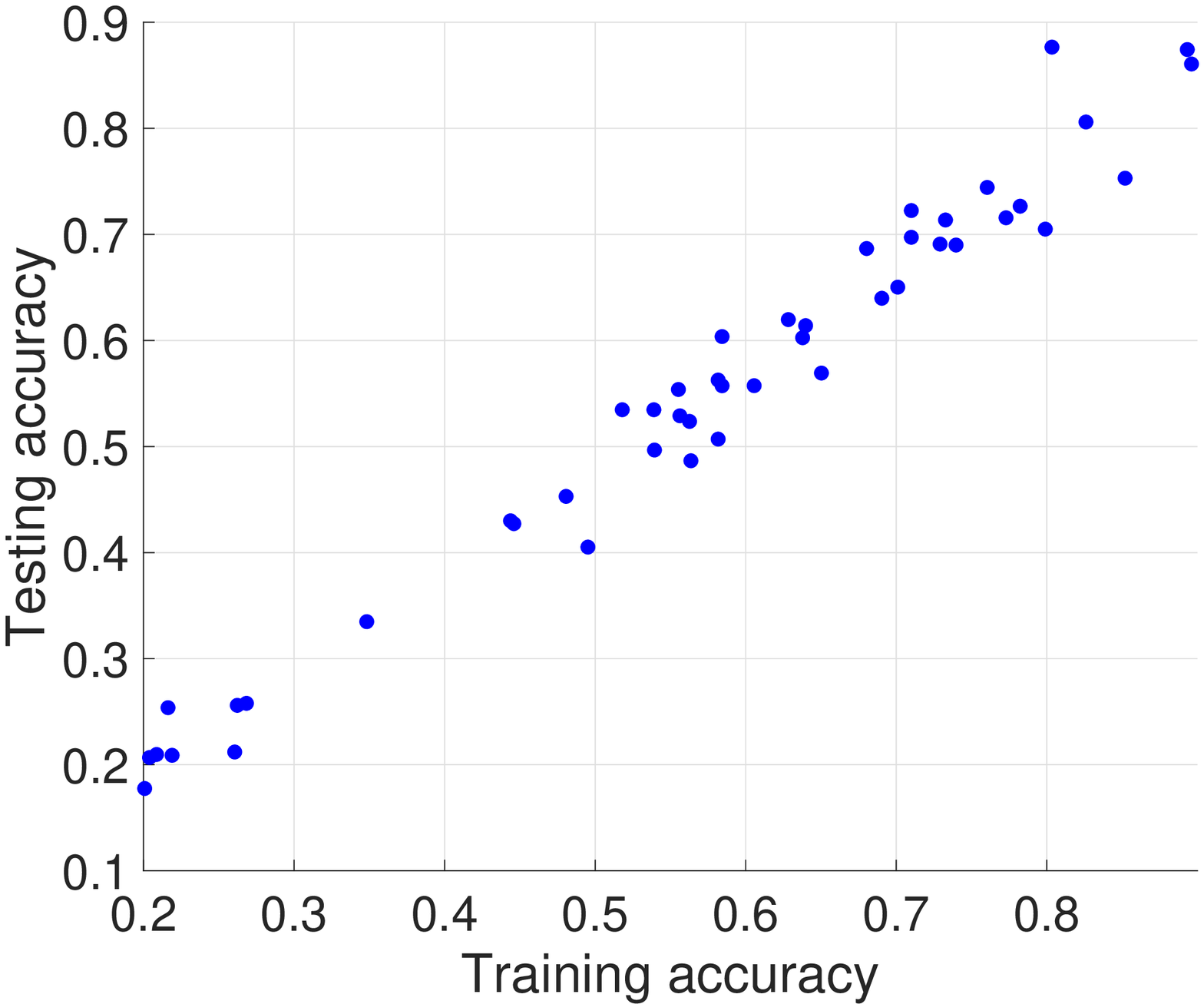}
    \caption{The relationship between training accuracy and testing accuracy.}
    \label{fig11}
  \end{minipage}
  \vspace*{-5mm}

\end{figure*}

Next, we examine the global accuracy loss with the number of communication rounds when fixed local accuracy loss $\epsilon_l=0.2$. In this simulation, we still set $\beta_\text{up}=-15$dB and $\beta_\text{down}=15$dB. Fig. 10 shows the global model accuracy loss changes with the number of communication rounds. From Fig. 10, we can see that the global model accuracy loss decreases with the number of communication rounds. Moreover, the difference between the actual global accuracy loss and $\epsilon_l=0.2$ is always within $0.1$ when the learning convergences. Please note that SINR and SNR practically affect the global aggregation and local training respectively. 


Next, we examine whether the well trained model is effective for the test datasets. In this simulation experiment, the test datasets are drawn from the same distribution as the training data. We randomly select 3 UEs and calculate the testing accuracy every 2 communication rounds. As shown in Fig. 11, we can see that the testing accuracy increases with training accuracy, where $\epsilon_l$ is the local training accuracy and $\hat{\epsilon}_l$ is the testing training accuracy. We can also see, in general, the difference between the training accuracy and the testing accuracy is within $[0, 0.1]$.

\subsubsection{Examining the trade-off between the computing resources and communication resources under FL framework}
First, we examine the relationship between the global model accuracy and available bandwidth in the uplink. In this simulation experiment, we first assume that bandwidth in the downlink and computing resources are sufficient, and then we fix the required local model accuracy ($0.1, 0.2, 0.3$) to verify the relationship between the global model accuracy and the amount of available bandwidth in the uplink. From Fig. 12, we can see that the global model accuracy sharply increases in the beginning and then increases slowly with the amount of available bandwidth in the uplink, as the number of local UEs that can participate in global aggregation quickly increases with the amount of bandwidth in the beginning. When the amount of bandwidth increases to be sufficient, it has little effect on the transmission success probability of local models, and thus the global model accuracy keeps fairly steady. Moreover, in the beginning, higher local model accuracy (lower $\epsilon_l$) leads to a higher global model accuracy. Specifically, the global model accuracy when $\epsilon_l=0.1$ is $8.5\%$ higher than that when $\epsilon_l=0.2$, while the global model accuracy when $\epsilon_l=0.2$ is $17.2\%$ higher than that when $\epsilon_l=0.3$.

\begin{figure*}[htbp]
\vspace*{-6mm}
  \centering
  \begin{minipage}{0.32\linewidth}
    \includegraphics[width=0.9\linewidth]{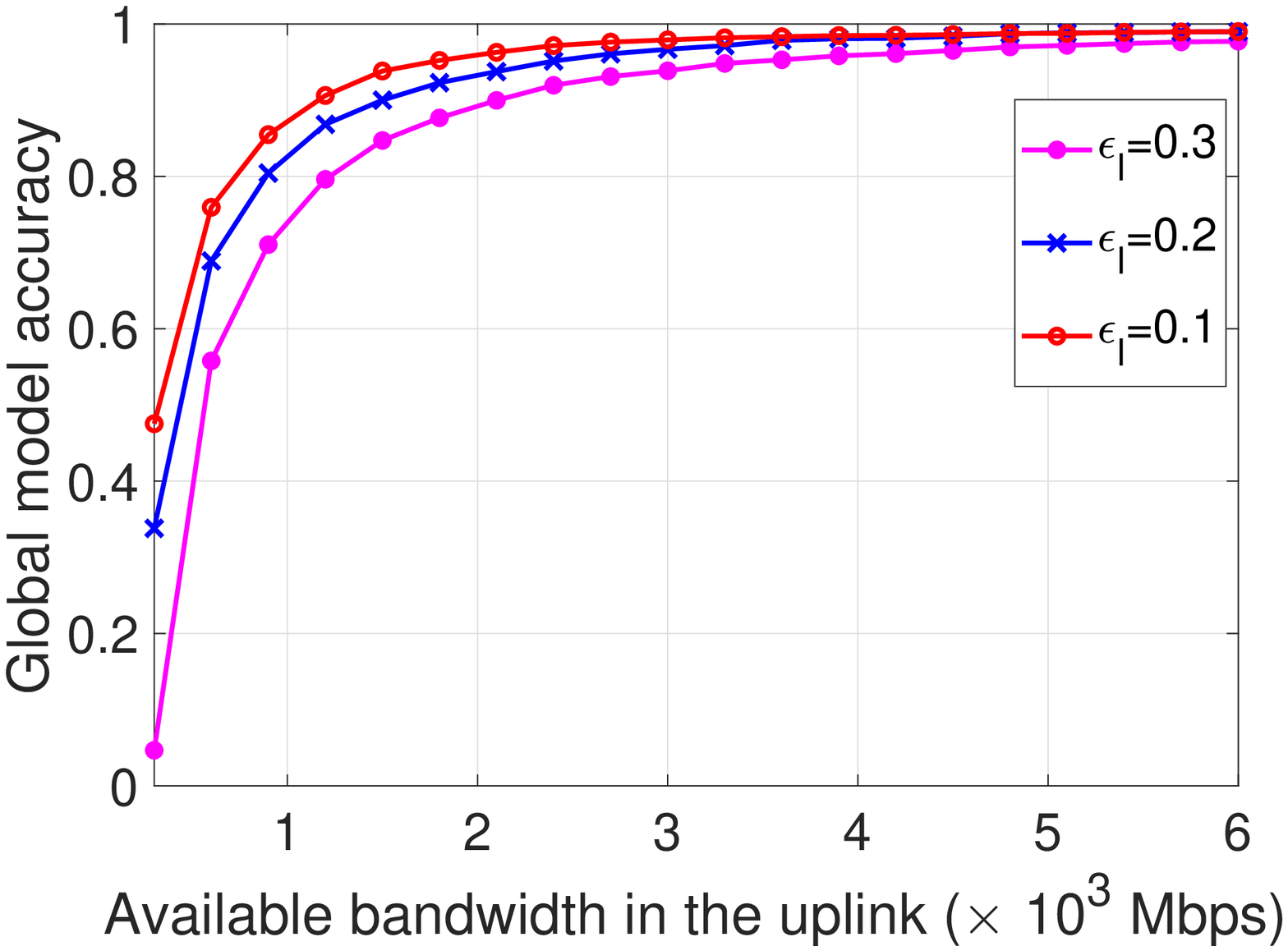}
    \caption{The relationship between available bandwidth in the uplink and global model accuracy.}
    \label{fig12}
  \end{minipage}
  \begin{minipage}{0.32\linewidth}
    \centering
    \includegraphics[width=0.9\linewidth]{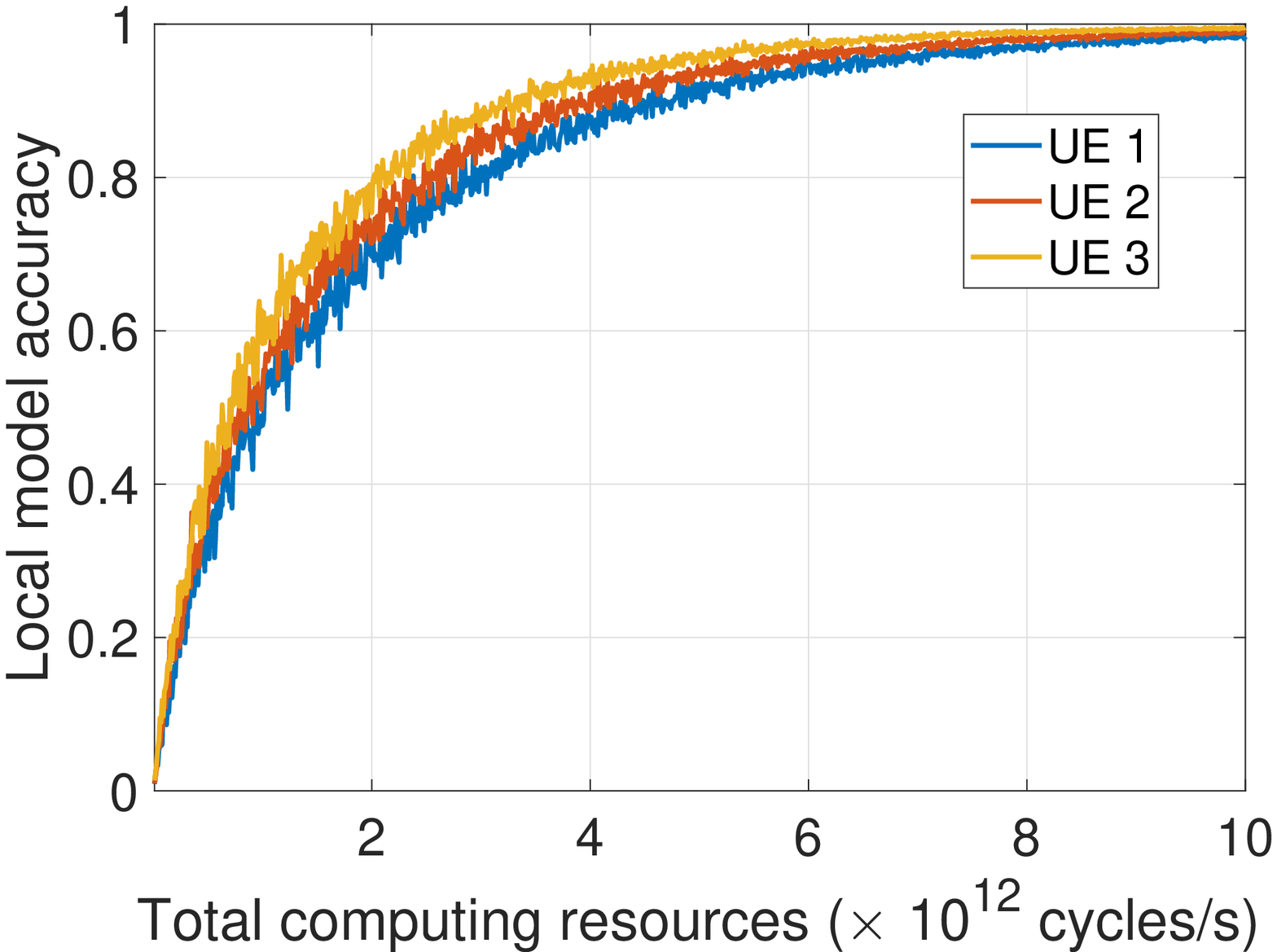}
    \caption{The relationship between available computing resources and local model accuracy.}
    \label{fig13}
  \end{minipage}
  \begin{minipage}{0.32\linewidth}
    \centering
    \vspace{-4mm}
    \includegraphics[width=0.9\linewidth]{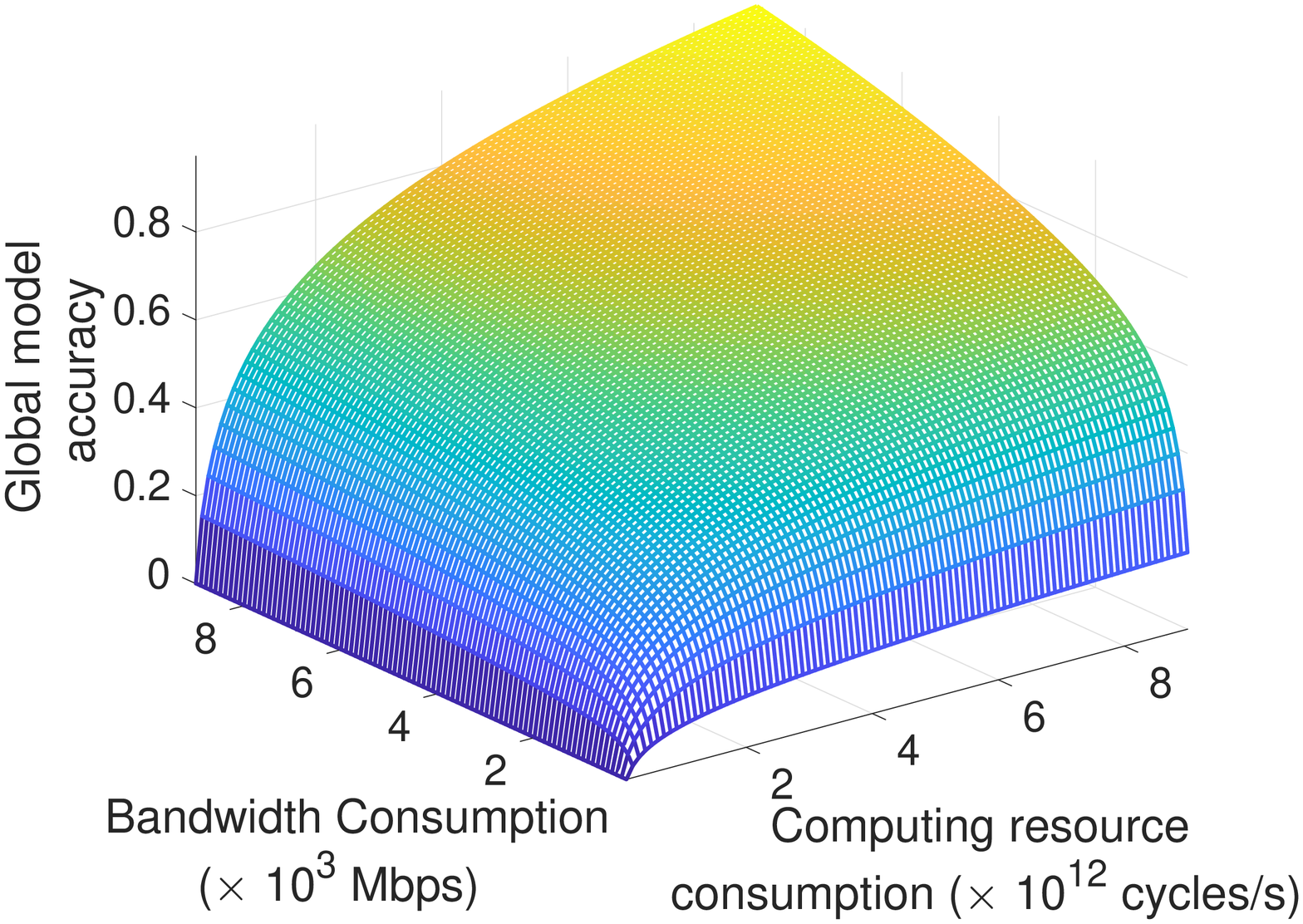}
    \caption{The trade-off between the computing resources and bandwidth.}
    \label{fig14}
  \end{minipage}
\vspace*{-5mm}
\end{figure*}



After that, we examine the relationship between the local model accuracy and computing resources, as shown in Fig. 13, where we randomly select 3 different UEs and assume that the bandwidth in the uplink/downlink is sufficient. From Fig. 13, we can see that the local model accuracy quickly increases in the beginning and then keeps fairly steady with the amount of computing resources. The reason is similar to that in Fig. 12, \emph{i.e.}, more computing resources leads to more local iterations in the beginning, while sufficient computing resources have little effect on local iterations. Please note the fluctuations of the curves in the Fig. 13 are due to that the local model accuracy in the Fig. 13 is recorded per local iteration, while the global accuracy in the Fig. 12 is recorded per communication round composed of 30 local iterations.

Finally, we examine the trend of the global model accuracy with respect to the amount of computing resources and the amount of bandwidth. Fig. 14 shows the relationship among the global model accuracy, computing resources, and bandwidth, where the trade-off between the amount of computing resources and the amount of bandwidth is verified. As shown in Fig. 14, both the amount of bandwidth and the amount of computing resources affect the global model accuracy, where we can flexibly adjust the amount of computing resources and the amount of bandwidth to guarantee a specific global model accuracy. Specifically, when we fix the amount of bandwidth used for transmitting the local models, we can increase the global model accuracy by increasing the amount of computing resources. When we fix the amount of computing resources used for local training, we can increase the global model accuracy by increasing the amount of bandwidth.

\section{Conclusion}
Wireless edge network intelligence enabled by FL has been widely acknowledged as a very promising means to address a wide range of challenging network issues. In this paper, we have theoretically analyzed how accurate of an ML model can be achieved by using FL and how many resources are consumed to guarantee a certain local/global accuracy. Specifically, we have derived the explicit expression of the model accuracy under FL framework, as a function of the amount of computing resources/communication resources for FL empowered wireless edge networks. Numerical results validate the effectiveness of our theoretical modeling. The modeling and results can provide some fundamental understanding for the trade-off between the learning performance and consumed resources, which is useful for promoting FL empowered wireless network edge intelligence. 


\appendices

\section{Calculate $\mu_I$ and $\sigma_I$}
\begin{small}
\begin{equation}\notag
\begin{split}
\setlength{\abovedisplayskip}{2pt}
\setlength{\belowdisplayskip}{2pt}
\mu_I=\overline{n}_IE(I_i)=\overline{n}_I \int_{d_2^{(i)}=d_\text{min}}^{d_0}P_{\text{up}}g(d_2^{(i)})f_{D_2^{(i)}}(d_2^{(i)})d(d_2^{(i)})
=\overline{n}_I \int_{d_2^{(i)}=d_\text{min}}^{d_0}P_{\text{up}}g(d_2^{(i)})\cdot \frac{2d_2^{(i)}}{(d_0)^2}d(d_2^{(i)})\\
=\frac{2P_{\text{up}}\overline{n}_I}{(d_0)^2}\int_{d_2^{(i)}=d_\text{min}}^{d_0} g(d_2^{(i)}) (d_2^{(i)})d(d_2^{(i)})
\overset{(b)}=\frac{2P_{\text{up}}n_I}{(d_0)^2}\int_{d_2^{(i)}=d_\text{min}}^{d_0} (d_2^{(i)}) dG(d_2^{(i)})~~~~~~~~~\\
\end{split}
\end{equation}
\end{small}
\begin{small}
\begin{equation}\notag
\begin{split}
\setlength{\abovedisplayskip}{2pt}
\setlength{\belowdisplayskip}{2pt}
=\frac{2P_{\text{up}}\overline{n}_I}{(d_0)^2}[(d_2^{(i)})G(d_2^{(i)})|_{d_2^{(i)}=d_\text{min}}^{d_0}-\int_{d_2^{(i)}=d_\text{min}}^{d_0} G(d_2^{(i)}) d(d_2^{(i)})],~~~~~~~
\end{split}
\end{equation}
\end{small}
where $\int_{d_2^{(i)}=d_\text{min}}^{d_0} G(d_2^{(i)})d(d_2^{(i)})={G}_2(d_0)-{G}_2(d_\text{min})$,
Therefore, $\mu_I$ can be given by
\begin{small}
\begin{equation}\notag
\setlength{\abovedisplayskip}{2pt}
\setlength{\belowdisplayskip}{2pt}
\mu_I=\frac{2P_{\text{up}}\overline{n}_I}{(d_0)^2}\{[d_0G(d_0)-d_\text{min}G(d_\text{min})]-
[{G}_2(d_0)-{G}_2(d_\text{min})]\},
\end{equation}
\end{small}
where $d_\text{min}$ is the minimum distance between UEs and BS, and we define $g(\cdot)=G'$ and $G(\cdot)={G}_2'(\cdot)$.
\begin{small}
\begin{equation}\notag
\setlength{\abovedisplayskip}{2pt}
\setlength{\belowdisplayskip}{2pt}
\sigma_I=\sqrt{\overline{n}_I}D(I_i)=\sqrt{\overline{n}_I}[E(I_i^2)-E^2(I_i)]
=\sqrt{\overline{n}_I}\left[\int_{d_2^{(i)}=d_\text{min}}^{d_0} 2P_{\text{up}}^2g^2(d_2^{(i)})\frac{d_2^{(i)}}{(d_0)^2}d(d_2^{(i)})-(\frac{\mu_I}{n_I})^2\right].
\end{equation}
\end{small}

\section{Proof of Proposition 2}

First, based on the definition of $G_i(\cdot)$ in problem (20), we have $\nabla^2 G_i(\cdot)=\nabla^2 F_i(\cdot)$.
Therefore, $\nabla^2 G_i(\cdot)$ also meets the $L$-smooth, $\gamma$-convex, and twice-differentiable assumptions, \emph{i.e.},
\vspace{-5mm}
\begingroup\makeatletter\def\f@size{10}\check@mathfonts
\def\maketag@@@#1{\hbox{\m@th\normalsize\normalfont#1}}
\begin{align}
\vspace{-5mm}
&\parallel\nabla G_i(w)-\nabla G_i(w')\parallel\leq L\parallel w-w'\parallel, \tag{B.1}\displaybreak[0]\\
&\parallel\nabla G_i(w)-\nabla G_i(w')\parallel\geq \gamma \parallel w-w'\parallel, \tag{B.2}\displaybreak[0]\\
&\gamma I\leq\nabla^2 G_i(w)\leq LI, \tag{B.3}\displaybreak[0]\\
&G_i(w)\geq G_i(w')+\nabla \left(G_i(w')\right)^T(w-w')+ \frac{\gamma}{2}\parallel w-w'\parallel^2.\tag{B.4}\displaybreak[0]
\end{align}
\endgroup

Then, when given $w_r$, we rewrite $G_i(\cdot)$ using the second-order Taylor expansion as
\begin{small}
\begin{equation}\notag
\begin{split}
\setlength{\abovedisplayskip}{2pt}
\setlength{\belowdisplayskip}{2pt}
G_i\left(w_r, h_i^{(r)(t+1)}\right)=G_i\left(w_r, h_i^{(r)(t)}\right)+
\left(h_i^{(r)(t+1)}-h_i^{(r)(t)}\right)^{\text{T}}
\cdot\nabla G_i\left(w_r, h_i^{(r)(t)}\right)+\\
\frac{1}{2}\left(h_i^{(r)(t+1)}-h_i^{(r)(t)}\right)^{\text{T}}\cdot\nabla^2 G_i\left(w_r, h_i^{(r)(t)}\right).
\cdot\left(h_i^{(r)(t+1)}-h_i^{(r)(t)}\right).
\end{split}
\end{equation}
\end{small}

As in GD method, we have 
\begin{small}
\begin{equation}\notag
\setlength{\abovedisplayskip}{2pt}
\setlength{\belowdisplayskip}{2pt}
h_i^{(r)(t+1)}=h_i^{(r)(t)}-\xi \nabla G_i(w_r, h_i^{(r)(t)}).
\end{equation}
\end{small}

Therefore, based on C.1 and C.3, we have
\begin{small}
\begin{equation}\notag
\begin{split}
\setlength{\abovedisplayskip}{2pt}
\setlength{\belowdisplayskip}{2pt}
G_i(w_r, h_i^{(r)(t+1)})=
G_i\left(w_r, h_i^{(r)(t)}\right)-\xi \parallel\nabla G_i\left(w_r, h_i^{(r)(t)}\right)\parallel^2+~~~~~~~~~~~~~~~~~~~~~~~~~~~~~~~~~~~~\\
\frac{\xi^2}{2}\parallel \nabla G_i\left(w_r, h_i^{(r)(t)}\right)\parallel^2\nabla^2 G_i\left(w_r, h_i^{(r)(t)}\right)~~~~~~~~~~~\\
\overset{(C.3)}{\leq} G_i\left(w_r, h_i^{(r)(t)}\right)-\xi \parallel\nabla G_i\left(w_r, h_i^{(r)(t)}\right)\parallel^2
+\frac{L\xi^2}{2}\parallel \nabla G_i\left(w_r, h_i^{(r)(t)}\right)\parallel^2\\
= G_i\left(w_r, h_i^{(r)(t)}\right)-
\frac{\left(2-L\xi\right)\xi}{2}\cdot \parallel \nabla G_i\left(w_r, h_i^{(r)(t)}\right)\parallel^2,~~~~~~~~~~~~~~~~~~~~~~
\end{split}
\end{equation}
\end{small}

~~~~Next we start to find the lower bound of $\parallel \nabla G_i\left(w_r, h_i^{(r)(t)}\right)\parallel^2$. For the optimal $h_i^{(r)^*}$ of $G_i(w_r, h_i^{(r)^*})$, we always have $\nabla G_i(w_r, h_i^{(r)^*})=0$. Therefore, we have
\begin{small}
\begin{equation}\tag{B.6}
\begin{split}
\setlength{\abovedisplayskip}{2pt}
\setlength{\belowdisplayskip}{2pt}
\parallel \nabla G_i\left(w_r, h_i^{(r)(t)}\right)\parallel^2=
\parallel \nabla G_i\left(w_r, h_i^{(r)(t)}\right)-\nabla G_i\left(w_r, h_i^{(r)^*}\right)\parallel^2~~~~~~~~~~~~~~~~~~~~~~~~~~~~~~~~~~\\
\overset{(C.2)}{\geq}\gamma \parallel \nabla G_i\left(w_r, h_i^{(r)(t)}\right)-\nabla G_i\left(w_r, h_i^{(r)^*}\right)\parallel
\parallel (w_r,h_i^{(r)(t)})-(w_r,h_i^{(r)^*})\parallel\\
\geq\gamma \left(\nabla G_i\left(w_r, h_i^{(r)(t)}\right)-\nabla G_i\left(w_r, h_i^{(r)^*}\right)\right)^{\text{T}}
\left( (w_r,h_i^{(r)(t)})-(w_r,h_i^{(r)^*})\right)\\
=\gamma \nabla G_i\left(w_r, h_i^{(r)(t)}\right)^{\text{T}}\left( (w_r,h_i^{(r)(t)})-(w_r,h_i^{(r)^*})\right)\\
\overset{(C.4)}{\geq} \gamma \left(G_i\left(w_r, h_i^{(r)(t)}\right)-G_i\left(w_r, h_i^{(r)^*}\right) \right).~~~~~~~~~
\end{split}
\end{equation}
\end{small}

Therefore, combining (B.5) and (B.6), we have
\begin{small}
\begin{equation}\notag
\begin{split}
\setlength{\abovedisplayskip}{2pt}
\setlength{\belowdisplayskip}{2pt}
 G_i\left(w_r, h_i^{(r)(t+1)}\right)-G_i\left(w_r, h_i^{(r)^*}\right) 
  \leq G_i\left(w_r, h_i^{(r)(t)}\right)-G_i\left(w_r, h_i^{(r)^*}\right)-~~~~~~~~~~~~~~~~~~~~~~~~~~~~~~~~~~~~~~~~~~~~~~\\
  \frac{\left(2-L\xi\right)\xi\gamma}{2}\left(G_i\left(w_r, h_i^{(r)(t)}\right)-G_i\left(w_r, h_i^{(r)^*}\right) \right)
 \overset{(\text{c})}{\leq} \left(1-\frac{\left(2-L\xi\right)\xi\gamma}{2}\right)^{(\tau+1)} \cdot
 \left(G_i\left(w_r, h_i^{(r)(0)}\right)-G_i\left(w_r, h_i^{(r)^*}\right) \right)\\
 \leq \text{exp}\left(-(\tau+1)\frac{\left(2-L\xi\right)\xi\gamma}{2}\right)
 \left(G_i\left(w_r, h_i^{(r)(0)}\right)-G_i\left(w_r, h_i^{(r)^*}\right) \right),
\end{split}
\end{equation}
\end{small}
where (c) can be obtained from $1-x\leq \text{exp}(-x)$. Therefore, to ensure that $G_i(w_r, h_i^{(r)(t)})-G_i(w_r, h_i^{(r)*})\leq
\epsilon_l(G_i(w_r, h_i^{(r)(0)})-G_i(w_r, h_i^{(r)*}))$, we have
\begin{equation}\notag
\setlength{\abovedisplayskip}{2pt}
\setlength{\belowdisplayskip}{2pt}
\epsilon_l\geq\text{exp}\left(-(\tau)\frac{\left(2-L\xi\right)\xi\gamma}{2}\right).
\end{equation}

Therefore, when $\xi <\frac{2}{L}$ (as $(2-L\xi)\xi\gamma>0$), we have
\begin{small}
\begin{equation}\notag
\setlength{\abovedisplayskip}{2pt}
\setlength{\belowdisplayskip}{2pt}
\tau\geq\frac{2}{(2-L\xi)\xi\gamma}\ln\left(\frac{1}{\epsilon_l}\right),
\end{equation}
\end{small}

\section{Proof of Proposition 3}
Under Assumption 1, Assumption 2, and Assumption 3, the following conditions hold on
\begin{small}
\begin{equation}\tag{C.1}
\setlength{\abovedisplayskip}{2pt}
\setlength{\belowdisplayskip}{2pt}
\frac{1}{L}\parallel\nabla F_i(w_r+h_i)-\nabla F_i(w_r)\parallel^2
\leq \left(\nabla F_i(w_r+h_i)-\nabla F_i(w_r)\right)^{\text{T}}h_i\\
\leq \frac{1}{\gamma}\parallel\nabla F_i(w_r+h_i)-\nabla F_i(w_r)\parallel^2,
\end{equation}
\end{small}
\begin{small}
\begin{equation}\tag{C.2}
\setlength{\abovedisplayskip}{2pt}
\setlength{\belowdisplayskip}{2pt}
\parallel\nabla F(w)\parallel^2\geq \gamma\left( F(w)-F(w^*)\right).
\end{equation}
\end{small}
~~~~The proof of (C.2) is similar to that of (B.6) in Appendix C. For (C.1), based on Lagrange median theorem, we always have a $w$ such that
\begin{small}
\begin{equation}\tag{C.3}
\setlength{\abovedisplayskip}{2pt}
\setlength{\belowdisplayskip}{2pt}
\nabla F_i(w_r+h_i)-\nabla F_i(w_r)=\nabla^2 F_i(w)h_i.
\end{equation}
\end{small}
For the optimal solution of local optimization problem, we always have  
\begin{small}
\begin{equation}\tag{C.4}
\setlength{\abovedisplayskip}{2pt}
\setlength{\belowdisplayskip}{2pt}
F_i\left(w_r+h_i^{(r)^*}\right)-\left(\nabla F_i(w_r)-\zeta\nabla F(w_r)\right)^{\text{T}}h_i^{(r)^*}=0
\end{equation}
\end{small}
~~~~With the above equalities and inequalities, we now start to prove Proposition 4.

\begin{small}
\begin{equation}\tag{C.5}
\begin{split}
\setlength{\abovedisplayskip}{2pt}
\setlength{\belowdisplayskip}{2pt}
F\left(w_{r+1}\left(\hat{\cal S}, \mathit{SINR}_{\text{up}}\right)\right)=F\left(\frac{\sum_{i=1}^n\hat{\cal S}_iw_i(r\tau)}{\sum_{i=1}^n\hat{\cal S}_i}\right)
=F\left(w_{r}\left(\hat{\cal S}, \mathit{SINR}_{\text{up}}\right)+\frac{\sum_{i=1}^n\hat{\cal S}_ih_i^{(r)(\tau)}}{\sum_{i=1}^n\hat{\cal S}_i}\right)~~~~~~~~~~~~~~~~~~~~~\\
\overset{(A1)}\leq F\left(w_{r}\left(\hat{\cal S}, \mathit{SINR}_{\text{up}}\right)\right)+
\frac{1}{\sum_{i=1}^n\hat{\cal S}_i}\sum_{i=1}^n\hat{\cal S}_i\nabla F\left(w_{r}\left(\hat{\cal S}, \mathit{SINR}_{\text{up}}\right)\right)^{\text{T}}h_i^{(r)(\tau)}+
+\frac{L}{2\left(\sum_{i=1}^n\hat{\cal S}_i\right)^2}\parallel\sum_{i=1}^n\hat{\cal S}_ih_i^{(r)(\tau)}\parallel^2.
\end{split}
\end{equation}
\end{small}
~~~~As
\vspace{-8mm}
\begingroup\makeatletter\def\f@size{10}\check@mathfonts
\def\maketag@@@#1{\hbox{\m@th\normalsize\normalfont#1}}
\begin{align}
\setlength{\abovedisplayskip}{2pt}
\setlength{\belowdisplayskip}{2pt}
&~~~~~~~~~~~~~~~~~~~~~~~~~G_i (w_r, h_i)=F_i(w_r+h_i)-(\nabla F_i(w_r)-\zeta\nabla F(w_r))^{\text{T}} h_i, \tag{C.6}\displaybreak[0]\\
&\parallel\frac{1}{\left(\sum_{i=1}^n\hat{\cal S}_i\right)^2}\sum_{i=1}^n\hat{\cal S}_ih_i^{(r)(\tau)}\parallel^2\leq
\left(\frac{1}{\sum_{i=1}^n\hat{\cal S}_i}\sum_{i=1}^n\parallel \hat{\cal S}_ih_i^{(r)(\tau)}\parallel\right)\parallel h_i^{(r)(\tau)}\parallel
\overset{\hat{\cal S}_i\geq0}\leq\frac{1}{\sum_{i=1}^n\hat{\cal S}_i}\sum_{i=1}^n\hat{\cal S}_i\parallel h_i^{(r)(\tau)}\parallel^2. \tag{C.7}\displaybreak[0]
\end{align}
\endgroup
~~~Therefore, we have
\begin{small}
\begin{equation}\notag
\begin{split}
\setlength{\abovedisplayskip}{2pt}
\setlength{\belowdisplayskip}{2pt}
F\left(w_{r+1}\left(\hat{\cal S}, \mathit{SINR}_{\text{up}}\right)\right)\leq F\left(g_{r}\left(\hat{\cal S}, \mathit{SINR}_{\text{up}}\right)\right)+
\frac{1}{\zeta\sum_{i=1}^n\hat{\cal S}_i}\sum_{i=1}^n\hat{\cal S}_i\{G_i\left(g_{r}\left(\hat{\cal S}, \mathit{SINR}_{\text{up}}\right), h_i^{(r)(\tau)}\right)-\\
F_i\left(w_{r}\left(\hat{\cal S}, \mathit{SINR}_{\text{up}}\right)+h_i^{(r)(\tau)}\right)+
\nabla F_i\left(w_{r}\left(\hat{\cal S}, \mathit{SINR}_{\text{up}}\right)\right)h_i^{(r)(\tau)}\}+
\frac{L}{2\left(\sum_{i=1}^n\hat{\cal S}_i\right)^2}\parallel\sum_{i=1}^n\hat{\cal S}_ih_i^{(r)(\tau)}\parallel^2\\
\overset{(A2)}\leq F\left(w_{r}\left(\hat{\cal S}, \mathit{SINR}_{\text{up}}\right)\right)+
\frac{1}{\zeta\sum_{i=1}^n\hat{\cal S}_i}\sum_{i=1}^n\hat{\cal S}_i\{G_i\left(w_{r}\left(\hat{\cal S}, \mathit{SINR}_{\text{up}}\right), h_i^{(r)(\tau)}\right)-~~~~~~~~~~~~~~~~~~~~~~~~~~\\
F_i\left(w_{r}\left(\hat{\cal S}, \mathit{SINR}_{\text{up}}\right)\right)-\frac{\gamma}{2}\parallel h_i^{(r)(\tau)}\parallel^2\}+
\frac{L}{2\left(\sum_{i=1}^n\hat{\cal S}_i\right)^2}\parallel\sum_{i=1}^n\hat{\cal S}_i h_i^{(r)(\tau)}\parallel^2~~~~~~~~~~~~~~~~\\
\overset{(C.7)}\leq F\left(w_{r}\left(\hat{\cal S}, \mathit{SINR}_{\text{up}}\right)\right)+
\frac{1}{\zeta\sum_{i=1}^n\hat{\cal S}_i}\sum_{i=1}^n\hat{\cal S}_i\{G_i\left(w_{r}\left(\hat{\cal S}, \mathit{SINR}_{\text{up}}\right), h_i^{(r)(\tau)}\right)-~~~~~~~~~~~~~~~~~~~~~~~~\\
F_i\left(w_{r}\left(\hat{\cal S}, \mathit{SINR}_{\text{up}}\right)\right)-\frac{\gamma-L\zeta}{2}\parallel h_i^{(r)(\tau)}\parallel^2\}\\
\overset{(C.6)}=F\left(w_{r}\left(\hat{\cal S}, \mathit{SINR}_{\text{up}}\right)\right)+
\frac{1}{\zeta\sum_{i=1}^n\hat{\cal S}_i}\sum_{i=1}^n\hat{\cal S}_i\{G_i\left(w_{r}\left(\hat{\cal S}, \mathit{SINR}_{\text{up}}\right), h_i^{(r)(\tau)}\right)-~~~~~~~~~~~~~~~~~~~~\\
G_i\left(w_{r}\left(\hat{\cal S}, \mathit{SINR}_{\text{up}}\right), h_i^{(r)^*}\right)-
G_i\left(w_{r}\left(\hat{\cal S}, \mathit{SINR}_{\text{up}}\right), 0\right)+\\
G_i\left(w_{r}\left(\hat{\cal S}, \mathit{SINR}_{\text{up}}\right), h_i^{(r)^*}\right)
-\frac{\gamma-L\zeta}{2}\parallel h_i^{(r)(\tau)}\parallel^2\}\\ 
\leq F\left(w_{r}\left(\hat{\cal S}, \mathit{SINR}_{\text{up}}\right)\right)-
\frac{1}{\zeta\sum_{i=1}^n\hat{\cal S}_i}\sum_{i=1}^n\hat{\cal S}_i\{\frac{\gamma-L\zeta}{2}\parallel h_i^{(r)(\tau)}\parallel^2+~~~~~~~~~~~~~~~~~~~~~~~~~\\
\left(1-\epsilon_l\right)[G_i\left(w_{r}\left(\hat{\cal S}, \mathit{SINR}_{\text{up}}\right), 0\right)-
G_i\left(w_{r}\left(\hat{\cal S}, \mathit{SINR}_{\text{up}}\right), h_i^{(r)^*}\right)]\}.
\end{split}
\end{equation}
\end{small}

According to equation (C.6), we can calculate $G_i\left(w_{r}, 0\right)$ and $G_i\left(w_{r}, h_i^{(r)^*}\right)$ as follows,
\begin{small}
\begin{equation}\notag
\setlength{\abovedisplayskip}{2pt}
\setlength{\belowdisplayskip}{2pt}
G_i\left(w_{r}\left(\hat{\cal S}, \mathit{SINR}_{\text{up}}\right), 0\right)=F_i\left(w_r(\hat{\cal S}, \mathit{SINR}_{\text{up}}\right),~~~~~~~~~~~~~~
\end{equation}
\end{small}
\begin{small}
\begin{equation}\notag
\begin{split}
\setlength{\abovedisplayskip}{2pt}
\setlength{\belowdisplayskip}{2pt}
G_i\left(w_{r}\left(\hat{\cal S}, \mathit{SINR}_{\text{up}}\right), h_i^{(r)^*}\right)=F_i\left( w_r\left(\hat{\cal S}, \mathit{SINR}_{\text{up}}\right)+h_i^{(r)^*}\right)\\
-[\nabla F_i \left(w_r\left(\hat{\cal S}, \mathit{SINR}_{\text{up}}\right)\right)-
\zeta\nabla F\left(w_r\left(\hat{\cal S}, \mathit{SINR}_{\text{up}}\right)\right)]^{\text{T}}h_i^{(r)^*}.
\end{split}
\end{equation}
\end{small}
Therefore, we have
\begin{small}
\begin{equation}\notag
\begin{split}
\setlength{\abovedisplayskip}{2pt}
\setlength{\belowdisplayskip}{2pt}
F\left(w_{r+1}\left(\hat{\cal S}, \mathit{SINR}_{\text{up}}\right)\right)\leq F\left(w_{r}\left(\hat{\cal S}, \mathit{SINR}_{\text{up}}\right)\right)-
\frac{1}{\zeta\sum_{i=1}^n\hat{\cal S}_i}\sum_{i=1}^n\hat{\cal S}_i\{\frac{\gamma-L\zeta}{2}\parallel h_i^{(r)(\tau)}\parallel^2+~~~~~~~~~~~~~~~~~~\\
\left(1-\epsilon_l\right)\{F_i\left(w_r(\hat{\cal S}, \mathit{SINR}_{\text{up}}\right)-
F_i\left( w_r\left(\hat{\cal S}, \mathit{SINR}_{\text{up}}\right)+h_i^{(r)^*}\right)+
[\nabla F_i \left(w_r\left(\hat{\cal S}, \mathit{SINR}_{\text{up}}\right)\right)-~~~~~~~~~~~~~~~~~\\
\zeta\nabla F\left(w_r\left(\hat{\cal S}, \mathit{SINR}_{\text{up}}\right)\right)]^{\text{T}}h_i^{(r)^*}\}\}~~~~~~~~~~~~~~\\
\overset{(C.4)}=F\left(w_{r}\left(\hat{\cal S}, \mathit{SINR}_{\text{up}}\right)\right)-
\frac{1}{\zeta\sum_{i=1}^n\hat{\cal S}_i}\sum_{i=1}^n\hat{\cal S}_i\{\frac{\gamma-L\zeta}{2}\parallel h_i^{(r)(\tau)}\parallel^2+
\left(1-\epsilon_l\right)\{F_i\left(w_r(\hat{\cal S}, \mathit{SINR}_{\text{up}}\right)-~~~~~\\
F_i\left( w_r\left(\hat{\cal S}, \mathit{SINR}_{\text{up}}\right)+h_i^{(r)^*}\right)+
\nabla F_i\left(w_r\left(\hat{\cal S}, \mathit{SINR}_{\text{up}}\right)+h_i^{(r)^*}\right)^{\text{T}}h_i^{(r)^*}\}~~~~~~~~~\\
\overset{A2}\leq F\left(w_{r}\left(\hat{\cal S}, \mathit{SINR}_{\text{up}}\right)\right)-
\frac{1}{\zeta\sum_{i=1}^n\hat{\cal S}_i}\sum_{i=1}^n\hat{\cal S}_i\{\frac{\gamma-L\zeta}{2}\parallel h_i^{(r)(\tau)}\parallel^2+
\frac{(1-\epsilon_l)\gamma}{2}\parallel h_i^{(r)^*}\parallel^2\}~~~~~~~~~~~~\\
\leq F\left(w_{r}\left(\hat{\cal S}, \mathit{SINR}_{\text{up}}\right)\right)-\frac{(1-\epsilon_l)\gamma}{2\zeta\sum_{i=1}^n\hat{\cal S}_i}\sum_{i=1}^n\hat{\cal S}_i\parallel h_i^{(r)^*}\parallel^2~~~~~~~~~~~~~~~~~~~~~~~~~~~~~~~~~~~~~~~~~~~~~~~~~~~~\\
\end{split}
\end{equation}
\end{small}
\begin{small}
\begin{equation}\notag
\begin{split}
\setlength{\abovedisplayskip}{2pt}
\setlength{\belowdisplayskip}{2pt}
\overset{A1}\leq F\left(w_{r}\left(\hat{\cal S}, \mathit{SINR}_{\text{up}}\right)\right)-
\frac{(1-\epsilon_l)\gamma}{2\zeta L^2\sum_{i=1}^n\hat{\cal S}_i}\sum_{i=1}^n\hat{\cal S}_i\parallel\nabla F_i\left(w_r\left(\hat{\cal S}, \mathit{SINR}_{\text{up}}\right)+h_i^{(r)^*}\right)~~~~~~~~~~~~~~~~~~~~~~~\\
-\nabla F_i\left(w_r\left(\hat{\cal S}, \mathit{SINR}_{\text{up}}\right)\right)\parallel^2~~~~~~~~~~~~\\
\overset{C.4}=F\left(w_{r}\left(\hat{\cal S}, \mathit{SINR}_{\text{up}}\right)\right)-
\frac{(1-\epsilon_l)\gamma\zeta}{2L^2\sum_{i=1}^n\hat{\cal S}_i}\sum_{i=1}^n\hat{\cal S}_i\parallel\nabla F\left(w_{r}\left(\hat{\cal S}, \mathit{SINR}_{\text{up}}\right)\right)\parallel^2~~~~~~~~~~~~~~~~~~~~~~~~~~~\\
\overset{C.2} \leq F\left(w_{r}\left(\hat{\cal S}, \mathit{SINR}_{\text{up}}\right)\right)-
\frac{(1-\epsilon_l)\gamma\zeta}{2L^2\sum_{i=1}^n\hat{\cal S}_i}\sum_{i=1}^n \hat{\cal S}_i\gamma\left(F\left(w_{r}\left(\hat{\cal S}, \mathit{SINR}_{\text{up}}\right)-F\left(w^*\right)\right)\right)~~~~~~~~~~~~~~~~\\
=F\left(w_{r}\left(\hat{\cal S}, \mathit{SINR}_{\text{up}}\right)\right)-
\frac{(1-\epsilon_l)\gamma^2\zeta}{2L^2}\left(F\left(w_{r}\left(\hat{\cal S}, \mathit{SINR}_{\text{up}}\right)-F\left(w^*\right)\right)\right).~~~~~~~~~~~~~~~~
\end{split}
\end{equation}
\end{small}

Therefore, based on (D.8), we have
\begin{small}
\begin{equation}\notag
\begin{split}
\setlength{\abovedisplayskip}{2pt}
\setlength{\belowdisplayskip}{2pt}
F\left(w_{r+1}\left(\hat{\cal S}, \mathit{SINR}_{\text{up}}\right)\right)-F\left(w^*\right)\leq
F\left(w_{r}\left(\hat{\cal S}, \mathit{SINR}_{\text{up}}\right)\right)-F\left(w^*\right)-~~~~~~~~~~~~~~~~~~~~~~~~~~~~~~~~~~\\
\frac{(1-\epsilon_l)\gamma^2\zeta}{2L^2}\left(F\left(w_{r}\left(\hat{\cal S}, \mathit{SINR}_{\text{up}}\right)\right)-F\left(w^*\right)\right)~~~~~~~~~~~~~~~~~~~~\\
=\left(1-\frac{(1-\epsilon_l)\gamma^2\zeta}{2L^2}\right) \left(F\left(w_{r}\left(\hat{\cal S}, \mathit{SINR}_{\text{up}}\right)\right)-F\left(w^*\right)\right)~~~~~~\\
\leq \left(1-\frac{(1-\epsilon_l)\gamma^2\zeta}{2L^2}\right)^{(r+1)} \left(F(w_0)-F\left(w^*\right)\right)
\overset{(\text{c})}\leq \text{exp}\left(-(r+1)\left(\frac{(1-\epsilon_l)\gamma^2\zeta}{2L^2}\right)\right)\left(F(w_0)-F\left(w^*\right)\right).
\end{split}
\end{equation}
\end{small}


To guarantee the global accuracy, \emph{i.e.}, $F(w_r)-F(w^*)\leq \epsilon_g ( F(w_0)-F(w^*))$, we have 
\begin{small}
\begin{equation}\notag
\setlength{\abovedisplayskip}{2pt}
\setlength{\belowdisplayskip}{2pt}
\epsilon_g\geq\text{exp}\left(-r\frac{(1-\epsilon_l)\gamma^2\zeta}{2L^2}\right).
\end{equation}
\end{small}
Therefore, when $0<\zeta<\frac{\gamma}{L}$, we have $r\geq \frac{2L^2\ln \frac{1}{\epsilon_g}}{(1-\epsilon_l)\gamma^2\zeta}$.


\bibliographystyle{IEEEtran}
\bibliography{reference}

\begin{thebibliography}{10}
\providecommand{\url}[1]{#1}
\csname url@samestyle\endcsname
\providecommand{\newblock}{\relax}
\providecommand{\bibinfo}[2]{#2}
\providecommand{\BIBentrySTDinterwordspacing}{\spaceskip=0pt\relax}
\providecommand{\BIBentryALTinterwordstretchfactor}{4}
\providecommand{\BIBentryALTinterwordspacing}{\spaceskip=\fontdimen2\font plus
\BIBentryALTinterwordstretchfactor\fontdimen3\font minus
  \fontdimen4\font\relax}
\providecommand{\BIBforeignlanguage}[2]{{%
\expandafter\ifx\csname l@#1\endcsname\relax
\typeout{** WARNING: IEEEtran.bst: No hyphenation pattern has been}%
\typeout{** loaded for the language `#1'. Using the pattern for}%
\typeout{** the default language instead.}%
\else
\language=\csname l@#1\endcsname
\fi
#2}}
\providecommand{\BIBdecl}{\relax}
\BIBdecl

\bibitem{IoT}
\BIBentryALTinterwordspacing
B.~Jovanovi{\'c}, ``{Internet of Things Statistics for 2021 -- Taking Things
  Apart}.'' [Online]. Available:
  \url{https://dataprot.net/statistics/iot-statistics/}
\BIBentrySTDinterwordspacing

\bibitem{lim2020federated}
W.~Y.~B. Lim, N.~C. Luong, D.~T. Hoang, Y.~Jiao, Y.-C. Liang, Q.~Yang,
  D.~Niyato, and C.~Miao, ``{Federated Learning in Mobile Edge Networks: A
  Comprehensive Survey},'' \emph{IEEE Communications Surveys \& Tutorials},
  vol.~22, no.~3, pp. 2031--2063, 2020.

\bibitem{yang2019federated}
Q.~Yang, Y.~Liu, T.~Chen, and Y.~Tong, ``{Federated Machine Learning: Concept
  and Applications},'' \emph{ACM Transactions on Intelligent Systems and
  Technology (TIST)}, vol.~10, no.~2, pp. 1--19, 2019.

\bibitem{jeong2018communication}
E.~Jeong, S.~Oh, H.~Kim, J.~Park, M.~Bennis, and S.-L. Kim,
  ``{Communication-efficient On-device Machine Learning: Federated Distillation
  and Augmentation under Non-iid Private Data},'' \emph{arXiv preprint
  arXiv:1811.11479}, 2018.

\bibitem{he2020group}
C.~He and M.~Annavaram, ``{Group Knowledge Transfer: Federated Learning of
  Large CNNs at the Edge},'' \emph{in proceedings of Advances in Neural
  Information Processing Systems 33 (NeurIPS 2020)}, no.~33, 2020.

\bibitem{chen2020convergence}
M.~Chen, H.~V. Poor, W.~Saad, and S.~Cui, ``{Convergence Time Optimization for
  Federated Learning over Wireless Networks},'' \emph{IEEE Transactions on
  Wireless Communications}, vol.~20, no.~4, pp. 2457--2471, 2020.

\bibitem{chen2020wireless}
{M. Chen,H. V. Poor, W. Saad, and S. Cui}, ``{Wireless Communications for
  Collaborative Federated Learning},'' \emph{IEEE Communications Magazine},
  vol.~58, no.~12, pp. 48--54, 2020.

\bibitem{9247530}
L.~U. Khan, S.~R. Pandey, N.~H. Tran, W.~Saad, Z.~Han, M.~N.~H. Nguyen, and
  C.~S. Hong, ``{Federated Learning for Edge Networks: Resource Optimization
  and Incentive Mechanism},'' \emph{IEEE Communications Magazine}, vol.~58,
  no.~10, pp. 88--93, 2020.

\bibitem{liu2020device}
Y.-J. Liu, G.~Feng, Y.~Sun, S.~Qin, and Y.-C. Liang, ``{Device Association for
  RAN Slicing based on Hybrid Federated Deep Reinforcement Learning},''
  \emph{IEEE Transactions on Vehicular Technology}, vol.~69, no.~12, pp.
  15\,731--15\,745, 2020.

\bibitem{9488679}
B.~Luo, X.~Li, S.~Wang, J.~Huang, and L.~Tassiulas, ``{Cost-Effective Federated
  Learning Design},'' in \emph{proceedings of IEEE INFOCOM 2021 - IEEE
  Conference on Computer Communications}, 2021, pp. 1--10.

\bibitem{8952884}
K.~Yang, T.~Jiang, Y.~Shi, and Z.~Ding, ``{Federated Learning via Over-the-Air
  Computation},'' \emph{IEEE Transactions on Wireless Communications}, vol.~19,
  no.~3, pp. 2022--2035, 2020.

\bibitem{9430906}
W.~Xia, W.~Wen, K.-K. Wong, T.~Q. Quek, J.~Zhang, and H.~Zhu,
  ``{Federated-Learning-Based Client Scheduling for Low-Latency Wireless
  Communications},'' \emph{IEEE Wireless Communications}, vol.~28, no.~2, pp.
  32--38, 2021.

\bibitem{9322580}
T.~Sery, N.~Shlezinger, K.~Cohen, and Y.~C. Eldar, ``{COTAF: Convergent
  Over-the-Air Federated Learning},'' pp. 1--6, 2020.

\bibitem{wen2021federated}
D.~Wen, K.-J. Jeon, and K.~Huang, ``{Federated Dropout--A Simple Approach for
  Enabling Federated Learning on Resource Constrained Devices},'' \emph{arXiv
  preprint arXiv:2109.15258}, 2021.

\bibitem{LAC}
C.~T. Dinh, N.~H. Tran, M.~N. Nguyen, C.~S. Hong, W.~Bao, A.~Y. Zomaya, and
  V.~Gramoli, ``{Federated Learning over Wireless Networks: Convergence
  Analysis and Resource Allocation},'' \emph{IEEE/ACM Transactions on
  Networking}, vol.~29, no.~1, pp. 398--409, 2020.

\bibitem{Ajoint}
M.~{Chen}, Z.~{Yang}, W.~{Saad}, C.~{Yin}, H.~V. {Poor}, and S.~{Cui}, ``{A
  Joint Learning and Communications Framework for Federated Learning Over
  Wireless Networks},'' \emph{IEEE Transactions on Wireless Communications},
  vol.~20, no.~1, pp. 269--283, 2021.

\bibitem{flint2017analysis}
I.~Flint, H.-B. Kong, N.~Privault, P.~Wang, and D.~Niyato, ``{Analysis of
  Heterogeneous Wireless Networks using Poisson Hard-core Hole Process},''
  \emph{IEEE Transactions on Wireless Communications}, vol.~16, no.~11, pp.
  7152--7167, 2017.

\bibitem{hunter2008transmission}
A.~M. Hunter, J.~G. Andrews, and S.~Weber, ``{Transmission Capacity of Ad Hoc
  Networks with Spatial Diversity},'' \emph{IEEE Transactions on Wireless
  Communications}, vol.~7, no.~12, pp. 5058--5071, 2008.

\bibitem{weber2010overview}
S.~Weber, J.~G. Andrews, and N.~Jindal, ``{An Overview of the Transmission
  Capacity of Wireless Networks},'' \emph{IEEE Transactions on Communications},
  vol.~58, no.~12, pp. 3593--3604, 2010.

\bibitem{SunBC}
Y.~{Sun}, L.~{Zhang}, G.~{Feng}, B.~{Yang}, B.~{Cao}, and M.~A. {Imran},
  ``{Blockchain-Enabled Wireless Internet of Things: Performance Analysis and
  Optimal Communication Node Deployment},'' \emph{IEEE Internet of Things
  Journal}, vol.~6, no.~3, pp. 5791--5802, 2019.

\bibitem{7110502}
Y.~J. Chun, M.~O. Hasna, and A.~Ghrayeb, ``Modeling heterogeneous cellular
  networks interference using poisson cluster processes,'' \emph{IEEE Journal
  on Selected Areas in Communications}, vol.~33, no.~10, pp. 2182--2195, 2015.

\bibitem{8340239}
V.~V. Chetlur and H.~S. Dhillon, ``{Coverage Analysis of a Vehicular Network
  Modeled as Cox Process Driven by Poisson Line Process},'' \emph{IEEE
  Transactions on Wireless Communications}, vol.~17, no.~7, pp. 4401--4416,
  2018.

\bibitem{hennig2007some}
C.~Hennig and M.~Kutlukaya, ``{Some Thoughts about the Design of Loss
  Functions},'' \emph{REVSTAT--Statistical Journal}, vol.~5, no.~1, pp. 19--39,
  2007.

\bibitem{yang2020energy}
Z.~Yang, M.~Chen, W.~Saad, C.~S. Hong, and M.~Shikh-Bahaei, ``{Energy Efficient
  Federated Learning over Wireless Communication Networks},'' \emph{IEEE
  Transactions on Wireless Communications}, vol.~20, no.~3, pp. 1935--1949,
  2020.

\bibitem{wang2021field}
J.~Wang, Z.~Charles, Z.~Xu, G.~Joshi, H.~B. McMahan, M.~Al-Shedivat, G.~Andrew,
  S.~Avestimehr, K.~Daly, D.~Data \emph{et~al.}, ``{A field Guide to Federated
  Optimization},'' \emph{arXiv preprint arXiv:2107.06917}, 2021.

\bibitem{Hsu1947}
P.~L. Hsu and H.~Robbins, ``{Complete Convergence and the Law of Large
  Numbers},'' \emph{in Proceedings of the National Academy of Sciences},
  vol.~33, no.~2, pp. 25--31, 1947.

\bibitem{DBLP}
\BIBentryALTinterwordspacing
T.~H. Hsu, H.~Qi, and M.~Brown, ``{Measuring the Effects of Non-Identical Data
  Distribution for Federated Visual Classification},'' \emph{CoRR}, vol.
  abs/1909.06335, 2019. [Online]. Available:
  \url{http://arxiv.org/abs/1909.06335}
\BIBentrySTDinterwordspacing

\bibitem{pmlr-v54-mcmahan17a}
\BIBentryALTinterwordspacing
B.~McMahan, E.~Moore, D.~Ramage, S.~Hampson, and B.~A. y~Arcas,
  ``{Communication-Efficient Learning of Deep Networks from Decentralized
  Data},'' \emph{in Proceedings of the 20th International Conference on
  Artificial Intelligence and Statistics}, vol.~54, pp. 1273--1282, 20--22 Apr
  2017. [Online]. Available:
  \url{http://proceedings.mlr.press/v54/mcmahan17a.html}
\BIBentrySTDinterwordspacing

\bibitem{gao2020end}
Y.~Gao, M.~Kim, S.~Abuadbba, Y.~Kim, C.~Thapa, K.~Kim, S.~A. Camtep, H.~Kim,
  and S.~Nepal, ``{End-to-End Evaluation of Federated Learning and Split
  Learning for Internet of Things},'' \emph{in Proceedings of 2020 IEEE
  Computer Society International Symposium on Reliable Distributed Systems
  (SRDS)}, pp. 91--100, 2020.

\bibitem{liu2020deep}
Y.~Liu, S.~Garg, J.~Nie, Y.~Zhang, Z.~Xiong, J.~Kang, and M.~S. Hossain,
  ``{Deep Anomaly Detection for Time-series Data in Industrial IoT: a
  Communication-efficient on-device Federated Learning Approach},'' \emph{IEEE
  Internet of Things Journal}, vol.~8, no.~8, pp. 6348--6358, 2020.

\bibitem{HHY}
H.~H. {Yang}, Z.~{Liu}, T.~Q.~S. {Quek}, and H.~V. {Poor}, ``{Scheduling
  Policies for Federated Learning in Wireless Networks},'' \emph{IEEE
  Transactions on Communications}, vol.~68, no.~1, pp. 317--333, 2020.

\bibitem{lecun1998mnist}
Y.~LeCun, ``{The MNIST Database of Handwritten Digits},'' \emph{http://yann.
  lecun. com/exdb/mnist/}, 1998.

\bibitem{li2017convergence}
Y.~Li and Y.~Yuan, ``{Convergence Analysis of Two-layer Neural Networks with
  ReLU Activation},'' in \emph{Proceedings of the 31st International Conference
  on Neural Information Processing Systems}, 2017, pp. 597--607.

\bibitem{krizhevsky2012imagenet}
A.~Krizhevsky, I.~Sutskever, and G.~E. Hinton, ``{Imagenet Classification with
  Deep Convolutional Neural Networks},'' \emph{Advances in neural information
  processing systems}, vol.~25, pp. 1097--1105, 2012.

\bibitem{darken1990note}
C.~Darken and J.~E. Moody, ``{Note on Learning Rate Schedules for Stochastic
  Optimization},'' in \emph{Proceedings of the 4th International Conference on
  Neural Information Processing Systems}, vol.~91, 1990, pp. 832--838.

\bibitem{liu2019variance}
L.~Liu, H.~Jiang, P.~He, W.~Chen, X.~Liu, J.~Gao, and J.~Han, ``{On the
  Variance of the Adaptive Learning Rate and Beyond},'' in \emph{Proceedings of
  International Conference on Learning Representations}, 2019.

\bibitem{liu2021access}
Y.-J. Liu, G.~Feng, J.~Wang, Y.~Sun, and S.~Qin, ``{Access Control for RAN
  Slicing based on Federated Deep Reinforcement Learning},'' \emph{in
  proceedings of ICC 2021-IEEE International Conference on Communications}, pp.
  1--6, 2021.

\end{thebibliography}

\end{document}